\documentclass{IEEEtaes}

\usepackage{color,array,amsthm}
\usepackage{graphicx}
\usepackage{subfigure}
\usepackage{cite}
\usepackage{amsmath,amssymb,amsfonts}
\usepackage{textcomp}
\usepackage{xcolor}
\usepackage{gensymb}
\usepackage{makecell}
\usepackage{stfloats}
\usepackage{textcomp}
\usepackage{multirow}
\usepackage{bm}
\usepackage{stfloats}
\usepackage{cuted}
\usepackage{algorithm}
\usepackage{mathrsfs}
\usepackage{algorithmic}

\begin{document}
\title{AWSPNet: Attention-based Dual-Tree Wavelet Scattering Prototypical Network for MIMO Radar Target Recognition and Jamming Suppression}

\author{Yizhen~Jia}
\author{Siyao~Xiao}
\affil{Tsinghua University, Shenzhen, China}
\author{Wenkai~Jia}
\author{Hui~Chen}
\member{Member, IEEE}
\author{Wen-Qin~Wang}
\member{Senior member, IEEE}
\affil{University of Electronic
	Science and Technology of China, Chengdu, China}

\receiveddate{Manuscript received 2025; revised XXXXX 00, 0000; accepted XXXXX 00, 0000.\\
This work was supported in part by the Postdoctoral Program for Innovation Talents under Grant BX20240054 and in part by the China Postdoctoral Science Foundation under Grant 2024M760355. \itshape (Corresponding author: Wen-Qin Wang)}
\authoraddress{Yizhen Jia, Wenkai Jia, Hui Chen, and Wen-Qin Wang are with School of Information and Communication Engineering, University of Electronic Science and Technology of China, Chengdu, 611731, P. R. China. Siyao Xiao is with the Shenzhen International Graduate School, Tsinghua University, Shenzhen 518055, Guangdong, China. (e-mail: Yizhen Jia: jiayizhen@std.uestc.edu.cn, Siyao Xiao: 202222230113@std.uestc.edu.cn, Wenkai Jia:mrwenkaij@126.com, Hui Chen: huichen0929@uestc.edu.cn, Wen-Qin Wang: wqwang@uestc.edu.cn)
}  

\maketitle
\begin{abstract}
	The increasing of digital radio frequency memory based electronic countermeasures poses a significant threat to the survivability and effectiveness of radar systems. These jammers can generate a multitude of deceptive false targets, overwhelming the radar's processing capabilities and masking targets. Consequently, the ability to robustly discriminate between true targets and complex jamming signals, especially in low signal-to-noise ratio (SNR) environments, is of importance. This paper introduces the attention-based dual-tree wavelet scattering prototypical network (AWSPNet), a deep learning framework designed for simultaneous radar target recognition and jamming suppression. The core of AWSPNet is the encoder that leverages the dual-tree complex wavelet transform to extract features that are inherently robust to noise and signal translations. These features are further refined by an attention mechanism and a pre-trained backbone network. To address the challenge of limited labeled data and enhance generalization, we employ a supervised contrastive learning strategy during the training phase. The classification is performed by a prototypical network, which is particularly effective in few-shot learning scenarios, enabling rapid adaptation to new signal types. We demonstrate the efficacy of our approach through extensive experiments. The results show that AWSPNet achieves 90.45\% accuracy at -6 dB SNR. Furthermore, we provide a physical interpretation of the network's inner workings through t-SNE visualizations, which analyze the feature separability at different stages of the model. Finally, by integrating AWSPNet with a time-domain sliding window approach, we present a complete algorithm capable of not only identifying but also effectively suppressing various types of jamming, thereby validating its potential for practical application in complex electromagnetic environments. The code is available in https://github.com/jiaxuanzhi/AwspNet.
\end{abstract}

\section{Introduction}
The operational effectiveness of modern radar systems is critically threatened by advanced electronic countermeasures, particularly those based on digital radio frequency memory (DRFM) technology \cite{zhangJoint2025,zhangJoint2024}. DRFM-based jammers can coherently intercept, manipulate, and retransmit radar signals to generate dense, realistic false targets that severely decrease the performance of radar\cite{grecoCombined2005, wangxuesongMathematic2007, LiuImpact2019}. This technological contest underscores the need for robust methodologies capable of accurate target discrimination and effective jamming suppression within contested electromagnetic environments \cite{rongqingIntegrated2024}. This paper is motivated by this challenge, focusing on the development of a unified framework designed to reliably distinguish between targets and deceptive jamming signals.

The task of jamming discrimination has been extensively investigated, with methodologies evolving from classical signal processing to modern deep learning approaches. Conventional methods predominantly relied on manually extracting discriminative features from various signal domains, such as time-frequency distributions, and then feeding them into machine learning classifiers like support vector machines or decision trees \cite{tianProduct2013}. For instance, \cite{ruihuiIntelligent2024} utilizes time-frequency features for the intelligent recognition of complex jamming. While effective to a degree, these methods often struggle with complex, agile, and compound jamming scenarios, as the handcrafted features may lack the robustness and generality required.

More recently, deep learning, particularly convolutional neural networks (CNNs), has become the dominant paradigm for jamming recognition due to its ability to learn hierarchical features automatically from raw or minimally processed data \cite{zhouRecognition2023}. Numerous studies have leveraged diverse CNN architectures, from foundational models to advanced networks like ResNet \cite{heDeep2016} and DenseNet \cite{huangDensely2017}, to classify jamming types from their time-frequency representations with high accuracy \cite{chenCompound2024, zhouCompound2024, lvMultilabel2024, quJRNet2020}. 
To address the inherent challenges of data dependency and noise sensitivity, recent research has explored more sophisticated approaches. For instance, \cite{liuPriorKnowledgeGuided2024} introduces a prior-knowledge-guided neural network that incorporates an interpretable scattering center layer, significantly improving radar target recognition with incomplete data and in low SNR environments. Similarly, \cite{wangApplication2024} utilizes a wavelet scattering network combined with ensemble learning for deception jamming recognition, a method that offers better interpretability than standard deep networks and maintains high performance even with minor sample sizes. Further tackling the issue of data scarcity, \cite{xiaoPSPNet2024} develops PSPNet, a prototypical network that employs pretraining and self-supervised fine-tuning to achieve robust, high-accuracy jamming recognition from very few labeled samples .
However, despite these advancements, several persistent challenges remain. Firstly, the {generalization problem} is a major concern; models trained on a specific dataset often exhibit a significant performance drop when faced with variations in signal parameters or low SNR conditions not seen during training \cite{shaDiffSwinT2023}. Secondly, the {sample problem} is prevalent, as collecting and labeling a large, diverse dataset of real-world jamming signals is often impractical, leading to few-shot learning challenges \cite{luoFewShot2023, luoFewshot2024, yangRadar2023}. This scarcity of data can lead to {overfitting}, where the model memorizes the training examples instead of learning generalizable features. Finally, the {interpretability problem} of deep neural networks, often referred to as the "black box" issue, makes it difficult to understand their decision-making process, hindering trust and adoption in critical defense applications \cite{wangApplication2024}.

To address the aforementioned limitations, this paper proposes a framework named the attention-based dual-tree wavelet scattering prototypical network (AWSPNet) for robust MIMO radar target recognition and jamming suppression. The AWSPNet architecture is designed to integrate the strengths of wavelet analysis, attention mechanisms, and metric-based few-shot learning. The pipeline begins by transforming the raw radar data into a robust feature representation using a dual-tree complex wavelet transform (DTCWT) module \cite{selesnickDualtree2005}. The DTCWT provides near shift-invariance and better directional selectivity compared to the standard discrete wavelet transform, yielding initial features that are less sensitive to signal translations and noise. Subsequently, an attention module is employed to dynamically re-weight these features, allowing the network to focus on the most informative aspects of the signal. The core feature extraction is performed by a pre-trained EfficientNetV2\cite{tanEfficientNetV22021} backbone, which is fine-tuned using a supervised contrastive learning objective \cite{khoslaSupervised2021}. This learning strategy explicitly encourages the encoder to produce an embedding space where samples from the same class are clustered closely together while samples from different classes are pushed far apart, significantly enhancing feature separability. Finally, instead of a conventional linear classifier, we utilize a prototypical network \cite{xiaoPSPNet2024} for classification. This metric-learning approach is exceptionally well-suited for few-shot scenarios, enhancing the model's ability to generalize to under-represented classes of signals.

The main contributions of this work are summarized as follows:
\begin{enumerate}
	\item We propose AWSPNet, which uses the DTCWT to extract jamming information from the radar's raw data, achieving robustness against noise and signal translations. The framework's generalization performance is further enhanced through a pre-training and fine-tuning strategy combined with a supervised contrastive learning architecture.
	\item We provide physical interpretability for the proposed network. By employing the t-SNE algorithm to visualize the feature distributions at key stages of the AWSPNet, we analyze the functional roles of its main modules and offer insights into how feature separability is progressively achieved.
	\item We design a practical target detection and jamming suppression algorithm based on the AWSPNet classifier. By applying a time-domain sliding window discrimination method, the system can effectively detect targets while concurrently identifying and suppressing various types of interference in the data stream.
\end{enumerate}
The rest of the paper is organized as follows: Section \ref{s2} introduces the MIMO model and problem formulation. Next, the method of AWSPNet is proposed in Section \ref{s3}. Numerical simulations are provided in Section \ref{s4}, and conclusions are finally drawn in Section \ref{s5}.

\textbf{\emph{Notations:}} $ \odot  $ represents the Hadamard product. $ * $ is the convolution operator. $ \left( \bullet \right) ^T $ and $ \left( \bullet \right) ^H $ denote the transpose and Hermitian transpose, respectively. $ \left( \bullet \right) ^* $ implies complex conjugation. $\mathbb{C} $ stands for the complex numbers, respectively. $ j $ represents the imaginary unit.
\section{Signal Model and Problem Formulation}
\label{s2}
\subsection{MIMO Signal Model}
\begin{figure}[htp]	
	\centering
	{\includegraphics[width=0.49\textwidth]{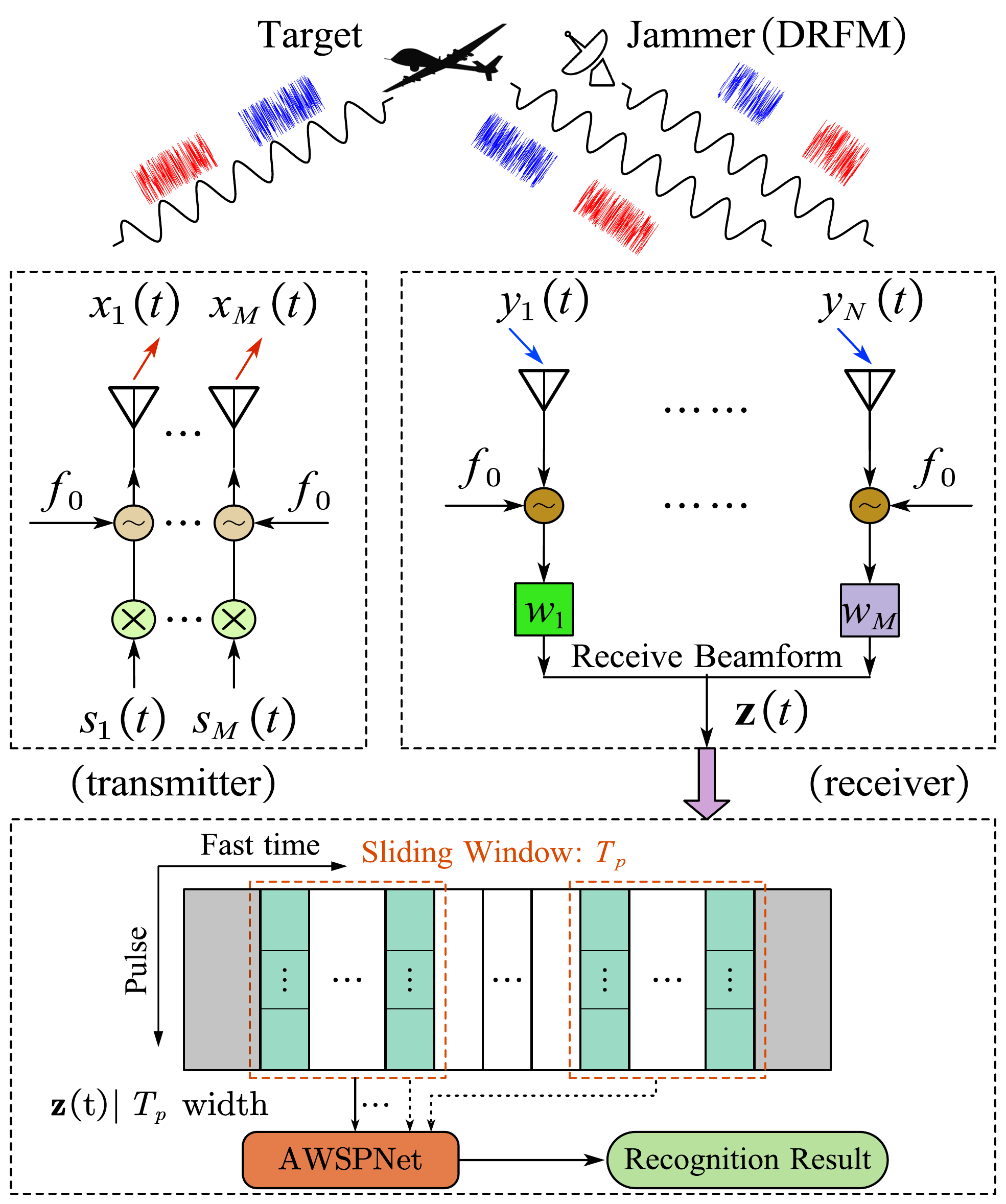}}
	\caption{ {General model of MIMO radar and target recognition processing with a target and jammers of DRFM.}}
	\label{f1}
\end{figure}
Consider a collocated MIMO radar system, it consists of $ M $ uniformly spaced transmitting elements and $ N $ receiving elements, as shown in Fig. \ref{f1}. It transmits $ Q $ pulses in a coherent processing interval (CPI). The $ m $-th transmit signal in one CPI, denoting by $x_m\left( t \right)  $, is that
\begin{equation}\label{1}
x_m\left( t \right) =s_m\left( t \right) \Pi  _{T_p}\left( t \right) e^{j2\pi f_0t}* \sum_{n=0}^Q{\delta \left( t-nT_r \right)},
\end{equation}
where $ s_m\left( t \right) ,m=1,\cdots,M $, is the baseband waveform of $ m $-th element. $ T_p $ is the pulse width, $ T_r $ is the pulse repeat interval (PRI). $ f_0 $ is the carrier frequency. $ \delta \left( t \right)  $ is Dirichlet function, and $ \Pi _{T_p}\left( t \right)  $ is the rectangle function which is defined as
\begin{equation}\label{2}
\Pi  _{T_p}\left( t \right) =\begin{cases}
		1,0<t<T_p\\
		0, others\\
	\end{cases}
\end{equation}
Let $\mathbf{y}\left( t \right)  \in \mathbb{C} ^{N\times 1} $ represents the echo signal vector of $ N $ receiving elements. i.e.,
\begin{equation}\label{3}
		\mathbf{y}\left( t \right) =\left[ y_1\left( t \right) ,\cdots ,y_N\left( t \right) \right] ^T
\end{equation}
Considering various active jammer from DRFM and clutter from the ground or sea, the echo can be represented by
\begin{equation}\label{4}
	\mathbf{y}\left( t \right) =\mathbf{y}_T\left( t \right) +\mathbf{y}_J\left( t \right) +\mathbf{y}_C\left( t \right) +\mathbf{n}\left( t \right) 
\end{equation}
where $\mathbf{y}_T\left( t \right) ,\mathbf{y}_J\left( t \right)$, and $ \mathbf{y}_C\left( t \right)   $ are the target, jammer and clutter echo respectively. $ \mathbf{n}\left( t \right)  $ represents the background noise which can be Gaussian or non-Gaussian noise. Because there are many kinds of active deception jamming, $ \mathbf{y}_J\left( t \right)  $ has no unified mathematical expression, so it is replaced by formal symbols in \eqref{4}, and more interference details are discussed in the next Section \ref{S13}. In this paper, we do not discuss the problem of clutter suppression, but consider the clutter as a kind of noise background, and investigate whether the interference type can be correctly identified in the clutter environment. 
Then, using receive beamforming, we obtain
\begin{equation}\label{key}
z\left( t \right) =\mathbf{w}_{r}^{H}\mathbf{y}\left( t \right) 
\end{equation}
where $ \mathbf{w}_r=\left[ w_1,\cdots ,w_N \right] ^T\in \mathbb{C} ^{N\times 1} $ is the receive beamforming weighted vector, usually it equals to the spatial steering vector pointing to the desired spatial area.

Fig. \ref{f1} illustrates the signal processing architecture proposed in this paper. Generally speaking, for the general time-domain orthogonal MIMO radar, in order to obtain the virtual array gain at the receiving end, it is necessary to use the matched filter to separate $ M $ transmit channels at each receiving channel. However, this process also destroys the time-frequency structure of the received echo, so that a large number of false targets are also presented as time-domain peaks similar to real targets after matched filtering, resulting in false alarms. In this sense, the matched filtering processing of time-domain orthogonal MIMO radar is not conducive to interference identification. 
Thus, unlike conventional radar processing pipelines, our procedure begins with the raw echo signal $ z\left( t \right) $ prior to pulse compression. It is the starting point of our proposed AWSPNet. Note that, $ z\left( t \right) $ contains abundant waveform information regarding both the target and potential interference, thereby aiding in their discrimination. To reduce the computational load on the subsequent AWSPNet, it may undergo receive beamforming to obtain $ z\left( t \right) $. 

\textbf{Remark:} Generally, for MIMO radar systems, if one intends to exploit the virtual $ M \times N $ multi-channel property to enhance beamforming performance, it is necessary to utilize the orthogonality among transmitted waveforms. This allows for the separation of transmit channels via matched filtering, albeit at the cost of losing time-domain waveform information. However, for certain specialized MIMO radars (e.g., frequency diverse array MIMO), time-domain waveform information is preserved through frequency-domain orthogonality. In such instances, virtual multi-channels can be obtained, enabling the use of the proposed AWSPNet for effective interference discrimination.
\subsection{Active Deception Jamming Echo Model}
\label{S13}
Recently, new types of deception jamming are emerged ceaselessly, mainly includes: 
\begin{itemize}
	\item \textbf{Dense false target category}:
	
	 1) dense false target jamming (DFTJ); 2) intermittent sampling and forwarding jamming (ISFJ)\cite{rongqingIntegrated2024}(in some literatures, it is also called chopping \& forwarding jamming) \cite{sparrowECM2006}; 3) intermittent sampling and repeat forwarding jamming (ISRJ) (in some literatures, it is also called chopping \& interleaving jamming) \cite{wangxuesongMathematic2007}; 4) smeared spectrum jamming (SSJ) \cite{sparrowECM2006}; 5) comb spectrum jamming (CSJ) \cite{wangInterference2023}.
	 
	 \item \textbf{Pull-off category}: 
	 
	 1) range gate pull off jamming (RGJ); 2) velocity gate pull off jamming (VGJ); 3) range-velocity gate pull off jamming (RVGJ)
\end{itemize}

To analyze the structure of interference signals, let $ \mathbf{a}_t\left( \theta_i  \right)   $ and $\mathbf{a}_r\left( \theta_i  \right) $ represent the transmit and receive spatial steering vector (SV) of the $ i $-th target, respectively. 
According to different radar array configurations, SV has different expressions (If it is a two-dimensional array, then the angle also has two dimensions). To simplify the expression, we assume that the array configuration is a one-dimensional uniformly spaced linear array (element spacing is half wavelength), then the SV can be written as:
\begin{equation}\label{key}
	\begin{aligned}
		\mathbf{a}_r\left( \theta _i \right) &=\left[ 1,e^{j\pi \sin \left( \theta _i \right)},\cdots ,e^{j\pi \left( N-1 \right) \sin \left( \theta _i \right)} \right] ^T\\
		\mathbf{a}_{t}^{}\left( \theta _i \right) &=\left[ 1,e^{j\pi \sin \left( \theta _i \right)},\cdots ,e^{j\pi \left( M-1 \right) \sin \left( \theta _i \right)} \right] ^T\\
	\end{aligned}
\end{equation}
Then, let $ r\left( t \right)  $ represents the signal received by DRFM (For simplicity, we assume that the DRFM employs a single antenna for signal reception), i.e.,
\begin{equation}\label{12}
r\left( t \right) \approx p_d\left( t \right) \mathbf{w}_{t}^{H}\left[ \mathbf{a}_{t}^{}\left( \theta \right) \odot \mathbf{x}\left( t-\tau \right) \right] ^T
\end{equation}
where $ \mathbf{w}_{t} $ is the transmit weighting, $ p_d\left( t \right) =e^{-j2\pi f_dt}$ is the Doppler frequency vector, and $f_d=2v_tf_0/c $ is the Doppler frequency. The angle and range (denoted by delay) of DRFM are $ \theta  $ and $ \tau $ respectively. Usually, In order to achieve the jamming effect, the angle of the jammer $ \theta $ is close to the target. Due to the existence of the deception mechanism, the real jammer delay parameter $ \tau $ is difficult to be reflected in the radar received echo. 

\textbf{Remark}: It is worth noting that the waveform emitted by the subsequent jammer is a re-sampled quantized waveform of \eqref{12}, and its spectral characteristics are different from the original spectrum. A large number of literature has been aligned for discussion, and interested readers can refer to\cite{grecoCombined2005}.
Therefore, the modeling of active deception jamming is based on the quantized signal $ 	r\left( t \right) $.
\subsubsection{DFTJ}
The main interference mechanism of DFTJ is to forward the signals in the DRFM storage unit multiple times in a short time, which can generate high-density coherent false targets at the radar receiving end to achieve the deception effect. 
DFTJ is divided into three types, including range-dense false target jamming (RDFJ), velocity-dense false target jamming (VDFJ), and range-velocity-dense false target jamming (RVDJ).
The modeling of RDFJ can be denoted by
\begin{equation}\label{key}
\mathbf{y}_{J}^{\left( \mathrm{RDFJ} \right)}\left( t \right) =\sum_{i=1}^{N_1}{\alpha _ir\left( t-\tau _i \right)}\mathbf{a}_r\left( \theta \right) 
\end{equation}
where $\alpha _i  $ is the modulated complex gain of $ i $-th false jammer by DRFM. $ \tau _i $ is the modulated delay of $ i $-th false jammer. The RDFJ makes multiple false targets appear in the same gate through appropriate delay forwarding, but it does not modulate the velocity information, so that the Doppler of these false targets coincides with the target Doppler. Similarly, VDFJ can also form a large number of false targets in the Doppler dimension.
\begin{equation}\label{key}
	\mathbf{y}_{J}^{\left( \mathrm{VDFJ} \right)}\left( t \right) =\mathbf{a}_r\left( \theta \right) \sum_{i=1}^{N_1}{\alpha _ir\left( t-\tau \right) e^{j2\pi f_it}}
\end{equation}
where $\tau  $ is the delay of jammer echo, $\alpha _i  $ is the modulated complex gain of $ i $-th VDFJ, and $f _i  $ is the modulated false Doppler frequency of $ i $-th VDFJ. In some literature, VDFJ is also called comb spectrum jamming.
Naturally, combining RDFJ and VDFJ, we obtain RVDJ
\begin{equation}\label{key}
	\mathbf{y}_{J}^{\left( \mathrm{RVDJ} \right)}\left( t \right) =\mathbf{a}_r\left( \theta \right) \sum_{i=1}^{N_1}{\alpha _ir\left( t-\tau _i \right) e^{j2\pi f_it}}
\end{equation}
RVDJ has the advantages of DFTJ in both distance and velocity dimensions, and the false target obtained by this modulation is closer to the real target in the actual environment. It is difficult for the receiver to judge the authenticity of the target, so as to achieve the purpose of deception.

\subsubsection{ISFJ and ISRJ}
\begin{figure}[htp]	
	\centering
	{\includegraphics[width=0.49\textwidth]{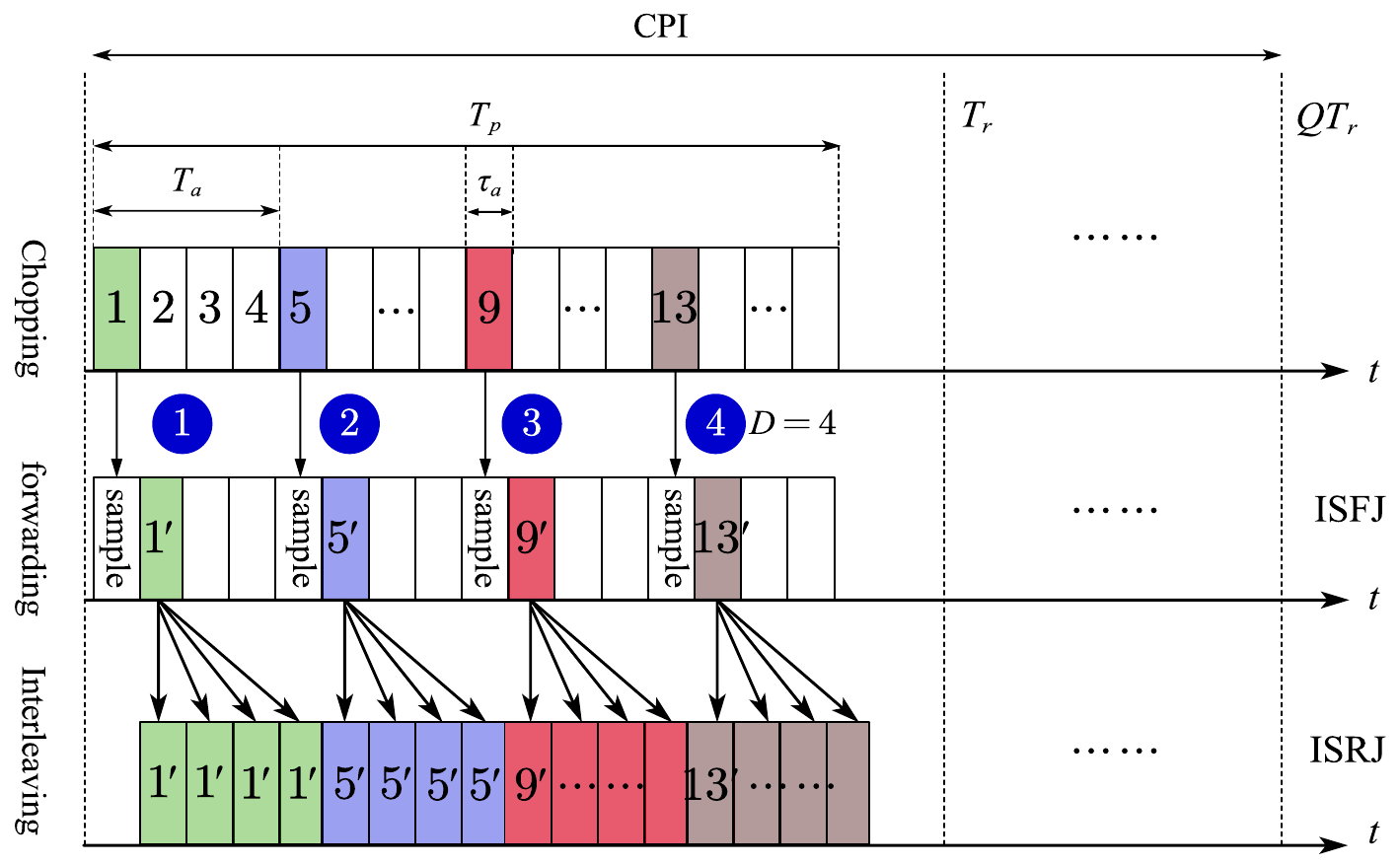}}
	\caption{ {Generate ISRJ and ISFJ.}}
	\label{f3}
\end{figure}

ISFJ and ISRJ are generated by a DRFM device. When the DRFM detector detects the rising edge of the pulse signal, it samples and stores the received signal for a period of time, and then modulates the stored slices according to the interference strategy and forwards them until the pulse falling edge is detected. The process is shown in Fig.\ref{f3}.
Let $ D $ represents the number of intermittent sampling for ISFJ or ISRJ, and $ T_a $ is the intermittent sampling interval. Then, we have $  D=\lfloor T_p/T_a \rfloor +1$, where $ \lfloor \cdot \rfloor  $ is take down the integer operator.
The modeling of ISFJ can be denoted by
\begin{equation}\label{key}
\begin{split}
	&\mathbf{y}_{J}^{\left( \mathrm{ISFJ} \right)}\left( t \right) =\mathbf{a}_r\left( \theta \right) r\left( t-\tau_c \right) \cdot 
\\
&\qquad \left[ \Pi _{\tau _a}\left( t \right) * \sum_{d=0}^D{\delta \left( t-dT_a \right)}* \sum_{n=0}^Q{\delta \left( t-nT_r \right)} \right] 
\end{split}
\end{equation}
Another, the modeling of ISRJ can be denoted by
\begin{equation}\label{key}
	\mathbf{y}_{J}^{\left( \mathrm{ISRJ} \right)}\left( t \right) =\mathbf{y}_{J}^{\left( \mathrm{ISFJ} \right)}\left( t \right) * \sum_{d=0}^D{\delta \left( t-d\tau _a \right)}
\end{equation}
Note that, $ \mathbf{y}_{J}^{\left( \mathrm{ISRJ} \right)}\left( t \right)$ and $\mathbf{y}_{J}^{\left( \mathrm{ISFJ} \right)}\left( t \right)  $ are the jammer signal received by the radar after an appropriate delay.
\subsubsection{SSJ}
\begin{figure}[htp]	
	\centering
	{\includegraphics[width=0.49\textwidth]{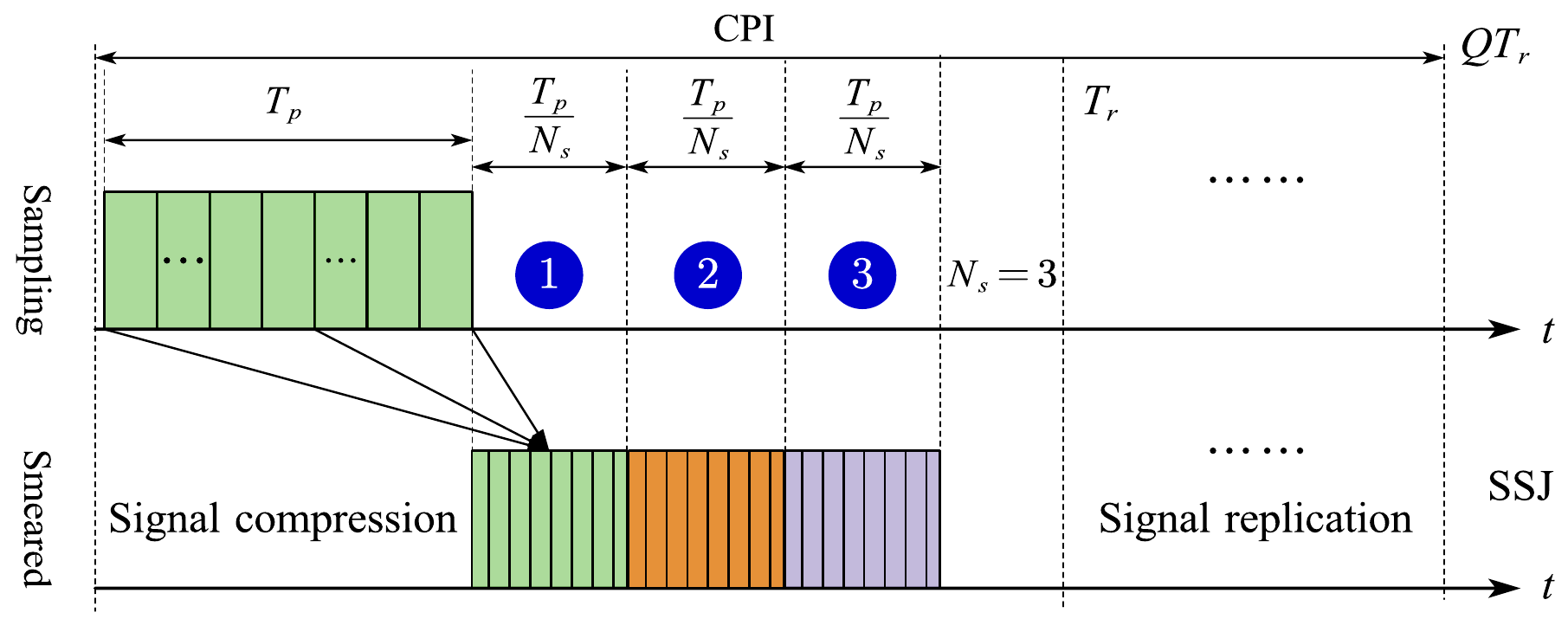}}
	\caption{ {Generate SSJ.}}
	\label{f4}
\end{figure}

The SSJ is composed of multiple sub-pulses, each of which is generated by interval sampling of radar signals. Therefore, false targets can be generated at the radar receiving end, which seriously affects the detection of real targets. The mathematical expression is
\begin{equation}\label{key}
	\mathbf{y}_{J}^{\left( \mathrm{SSJ} \right)}\left( t \right) =\mathbf{a}_r\left( \theta \right) \sum_{i=1}^{N_s}{\tilde{r}\left( t-i\frac{T_p}{N_s} \right)}
\end{equation}
where $ N_s $ is the sub-pulse number, and $ \tilde{r}\left( t \right)  $ is the compression version of $ r\left( t \right)  $. The essence of the smeared spectrum jamming is the compression and replication of the intercepted signal. The interference signal bandwidth is the same as the intercepted signal. Fig.\ref{f4} shows the generate process of SSJ.
\subsubsection{CSJ}
The CSJ is a superposition of a series of narrow-band interference signals within a specific bandwidth. It can be expressed as:
\begin{equation}\label{key}
	\mathbf{y}_{J}^{\left( \mathrm{CSJ} \right)}\left( t \right) =\mathbf{a}_r\left( \theta \right) \sum_{k=1}^K{\alpha _k} e^{j2\pi f_kt}
\end{equation}
where $ K $ is the number of comb teeth, $ \alpha _k $ is the complex amplitude of $ k $-th component, and $ f_k $ is the carrier frequency of $ k $-th component.
\subsubsection{Pull-off category jamming}
Pull-off category refers to the deceptive jamming signal that periodically separates the jamming signal from the real target parameter. Typical pull-off interference generally includes three periods: stand-off period, pull-off period and close period. Specifically, the automatic gain control (AGC) circuit captures the target and adjusts the gain. The true and false targets are gradually separated during the pull-in period. Due to the high energy and fast movement of the false target, the tracking system will track the false target and give up the true target. Finally, the transmitting signal radar loses the tracking target and returns to the search state during the closing period.

RGJ, VGJ, and RVGJ are based on the RDFJ, VDFJ, and RVDJ, receptively. Let $ P_f $ and $ P^* $ represent false and real target parameters, respectively.
\begin{equation}\label{key}
	\left\| P_f-P^* \right\| =\begin{cases}
		0,0\leqslant t<t_1,\mathrm{stand-off} \,\mathrm{period}\\
		0\rightarrow \delta , t_1\leqslant t<t_2,\mathrm{Pulling}\, \mathrm{period}\\
		P_f\,\mathrm{vanish}, t_2\leqslant t<T,\mathrm{Closing}\,\mathrm{period}\\
	\end{cases}
\end{equation}
During the stand-off period, $ P_f $ are essentially identical to $ P^* $, making it generally impossible for the radar to distinguish between the two. In the pulling period, $ P_f $ gradually diverge from those of the real target, and the pulling period remains within the radar’s tracking range. Consequently, the radar tracking system can receive the complete process of the false target’s parameter changes until the parameter deviation reaches the expected peak value $ \delta  $, entering the closing period. In the closing period, the deceptive interference signal disappears, and the radar tracking system is interrupted.

\section{Proposed AWSPNet}
\label{s3}
\begin{figure*}[htbp!]	
	\centering
	{\includegraphics[width=0.95\textwidth]{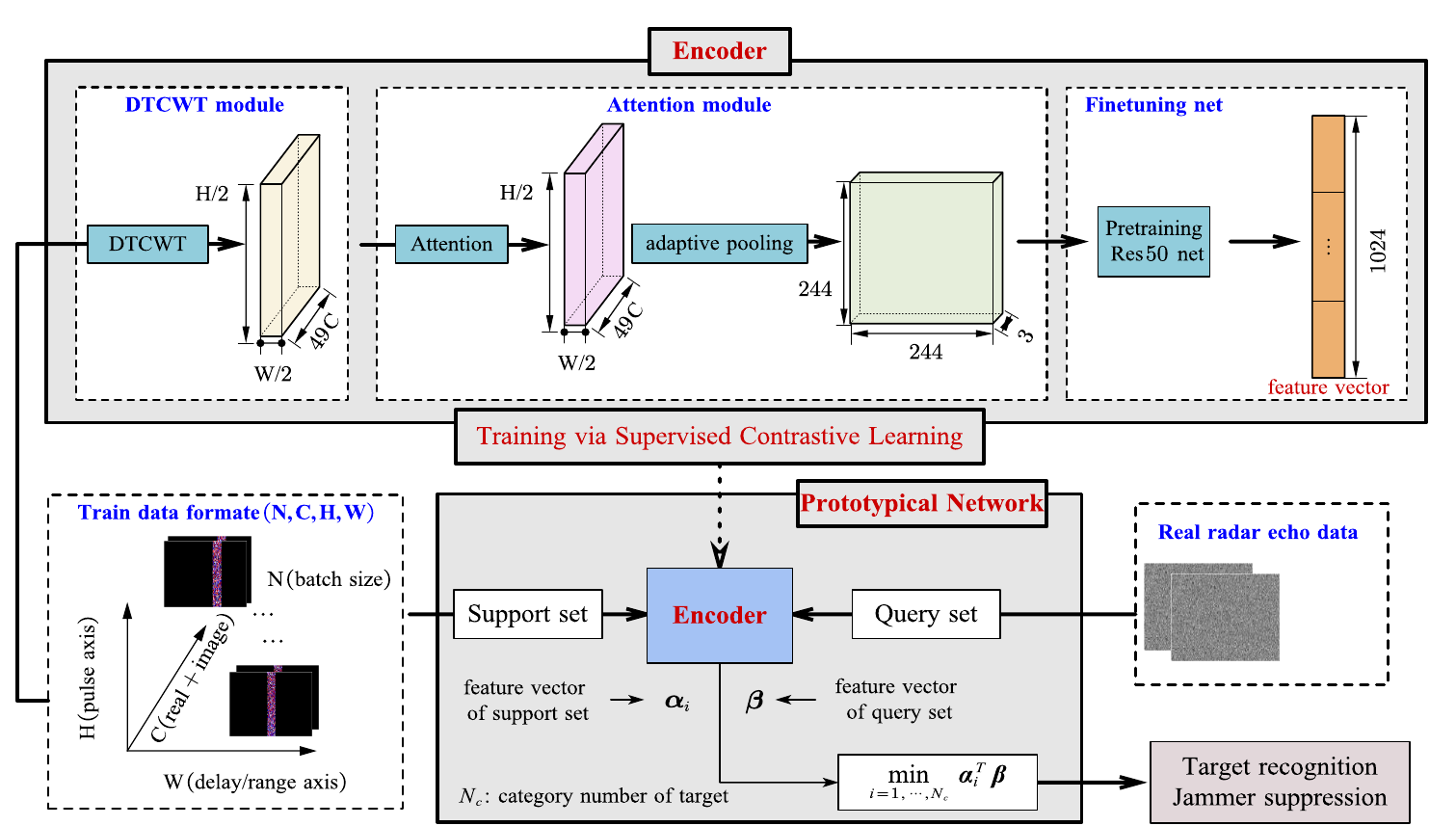}}
	\caption{ {Framework of the proposed AWSPNet that consists of two parts: encoder, and prototypical network. Encoder consists of DTCWT module, Attention module, and pre-training EfficientNet-based fine-tuning.}}
	\label{f4+}
\end{figure*}
The architecture of AWSPNet, illustrated in Fig \ref{f4+}, is strategically constructed to provide strong feature representation capabilities and enhanced generalization, particularly crucial for dynamic radar operational environments. The overall framework consists of two main stages: a sophisticated multi-component 'Encoder' for feature extraction from radar echo data, and a prototypical network leveraging these features for classification and few-shot learning.

The primary role of the 'Encoder' is to transform raw radar echo data, presented as a fast and slow time matrix, into highly discriminative feature vectors. This process begins with a DTCWT module, which leverages wavelet principles to extract initial features with inherent physical interpretability. These features are subsequently refined by an attention module designed to emphasize salient information. Finally, a pre-trained EfficientNet\cite{tanEfficientNetV22021} acts as a deep feature extractor, mapping the processed representations into a compact, high-dimensional feature vector. The entire Encoder is optimized through supervised contrastive learning, a strategy that aims to create a well-structured and discriminative embedding space.

Following feature extraction, the prototypical network is employed for the classification task. This network architecture is particularly advantageous for its efficacy in few-shot learning scenarios, thereby improving the system's adaptability to novel or sparsely represented radar signal classes. The prototypical network operates by learning class prototypes from a support set of examples within the feature space generated by the encoder. Query samples are then classified based on their similarity to these learned prototypes. This approach is selected over traditional classification networks to bolster generalization performance against unknown environmental conditions and diverse radar parameters, allowing for rapid adaptation in practical applications by defining local waveforms as the support set.

The synergy between the wavelet-based feature extraction, attention mechanisms, deep feature learning, and the prototypical classification framework enables AWSPNet to address the complexities of target recognition and jamming suppression in MIMO radar. Subsequent sections will provide a detailed exposition of each constituent module and the underlying methodologies.

\subsection{Dual-Tree Wavelet Scattering Network}
Dual-tree wavelet scattering network (DTWSN) is performed by the DTCWT which is an advanced signal and image processing tool over traditional wavelet transforms\cite{cotterUses2020}. Notably, DTCWT provides near shift-invariance (meaning the transform coefficients exhibit minimal changes with translations in the input signal) which contributes to its robustness against noise. Furthermore, DTCWT demonstrates stability in representing deformations, which is particularly beneficial for applications such as the stable characterization and detection of high-speed moving targets where signal or feature distortions are common. These characteristics make DTCWT a powerful method for feature extraction in dynamic and noisy environments.

\begin{figure}[htbp!]	
	\centering
	{\includegraphics[width=0.45\textwidth]{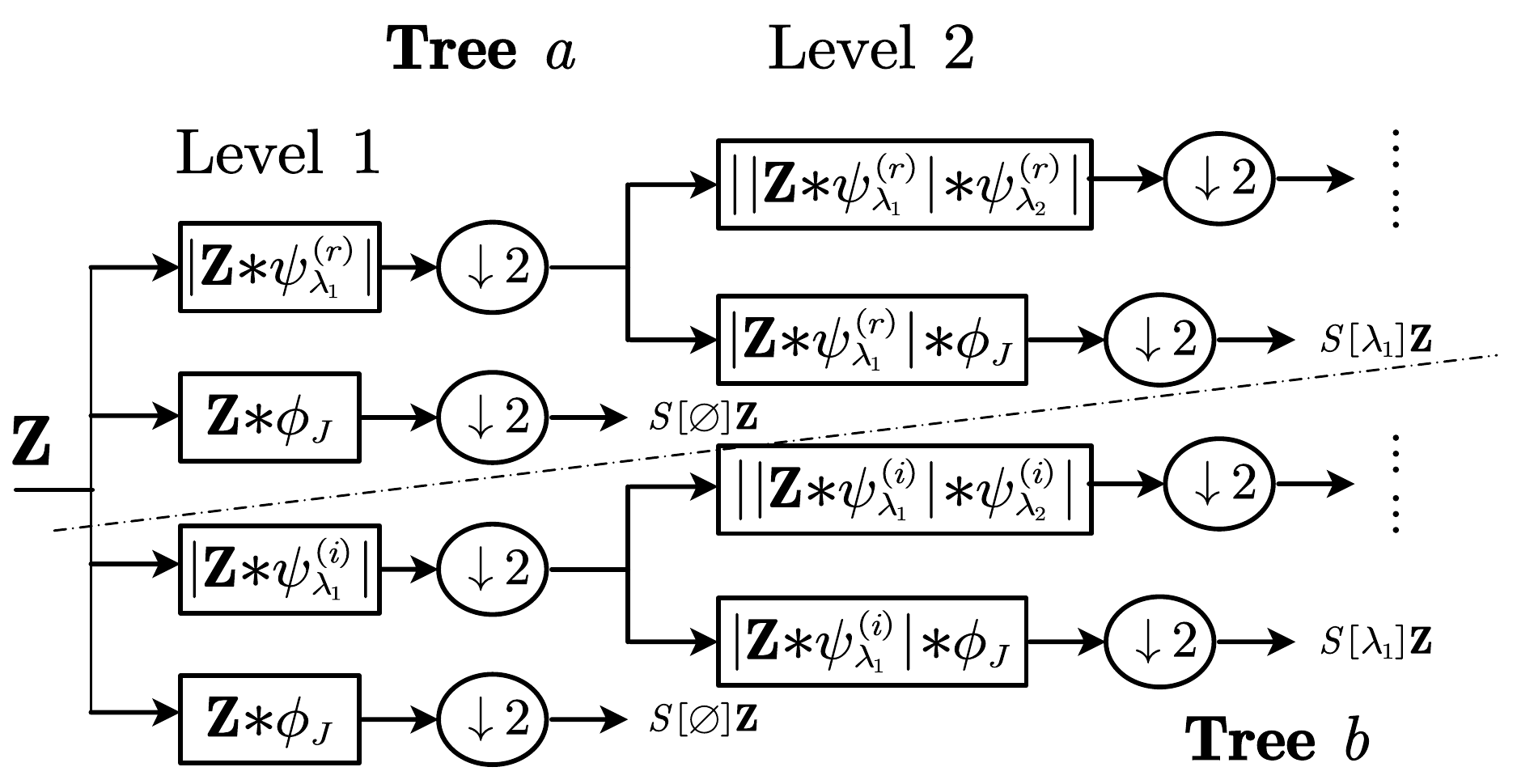}}
	\caption{ {Analysis of the flowchart for the 1-D DTCWT. Top 'tree' forms the real component of
			the complex wavelet $\psi _{\left( r \right)}  $, and the bottom 'tree' forms the imaginary (Hilbert pair) component $ \psi _{\left( i \right)} $.}}
	\label{f5}
\end{figure}

Consider a radar echo matrix $ \mathbf{Z}\in \mathbb{C} ^{Q\times L} $ after beamforming in one CPI, where $ L $ is the sampling number of fast time in one pulse. Then, we combines the wavelet transform and modulus operators into one operator $ \widetilde{\mathcal{W} } $, i.e.,
\begin{equation}\label{key}
\widetilde{\mathcal{W} }\mathbf{Z}=\left\{ \mathbf{Z}*\phi _J,\,\,\left| \mathbf{Z}*\psi _{\lambda} \right| \right\} _{\lambda}\triangleq \left\{ \mathbf{Z}*\phi _J, U[\lambda ]\mathbf{Z} \right\} _{\lambda}
\end{equation}
where $ \phi _J $ is the scaling function which is used to capture the low-frequency portion of the signal or an approximation thereof. $ \psi _{\lambda} $ is the wavelet function which is used to capture the high-frequency portion of the signal or its detailed components. $\lambda =(j,k)  $ with $ j\in \{1,2,...J\} $ indexing the scale, and $ k\in \{0,1,...K-1\} $ indexing the orientations of the chosen wavelet transform. $ * $ denotes the convolution operation. $ U[\lambda ]\mathbf{Z} $ is the input of the next decomposition layer, and we define
\begin{equation}\label{key}
	S\left[ \lambda \right] \mathbf{Z}=U[\lambda ]\mathbf{Z}*\phi _J=\left| \mathbf{Z}*\psi _{\lambda} \right|*\phi _J
\end{equation}
Then, $ S\left[ \lambda \right] \mathbf{Z} $ makes the first ordering scattering coefficients. Let $ p=\left( \lambda _1,\cdots ,\lambda _m \right)  $ be a path of length $ m $ describing the order of application of wavelets, then the $ m $-th layer wavelet scattering coefficient can be expressed as follows:
\begin{equation}\label{key}
	S\left[ p \right] \mathbf{Z}=\left| \left| \left| \mathbf{Z}*\psi _{\lambda _1} \right|*\psi _{\lambda _2} \right|\cdots \psi _{\lambda _m} \right|*\phi _J
\end{equation}
By stacking all the coefficients of $ p=\lambda _1,\left\{ \lambda _1,\lambda _2 \right\} ,\cdots \left\{ \lambda _1,\cdots ,\lambda _m \right\}  $, we obtain the extracted feature output of the scattering network, i.e.,
\begin{equation}\label{18}
	\mathbb{S} \left[ \cdot \right] \mathbf{Z}=\left\{ \begin{array}{c}
		\mathbf{Z}*\phi _J\\
		\left| \mathbf{Z}*\psi _{\lambda} \right|*\phi _J\\
		\left| \left| \mathbf{Z}*\psi _{\lambda _1} \right|*\psi _{\lambda _2} \right|*\phi _J\\
		\cdots\\
	\end{array} \right\} 
\end{equation}
Note that the above derivations apply to a general wavelet scattering network. In the case of the DTWSN, complex wavelets $ \tilde{\psi}={\psi}_{\left( r \right)}+j\cdot {\psi}_{\left( i \right)} $ are used (with their real part $ {\psi}_{\left( r \right)} $ and imaginary part $ {\psi}_{\left( i \right)} $ forming an orthogonal Hilbert pair). Consequently, the decomposition coefficients of the original signal in DTWSN consist of those from two trees: one from the real wavelet $ {\psi}_{\left( r \right)} $ and one from the imaginary wavelet $ {\psi}_{\left( i \right)} $ (with the coefficients for a single tree obtained from \eqref{18}), as shown in Fig \ref{f5}. For further derivation details, please refer to wavelet scattering network \cite{brunaInvariant2013}, DTCWT \cite{selesnickDualtree2005} and DTWSN \cite{cotterUses2020}.
The advantage of DTWSN is that it uses preselected wavelet filters instead of learning the coefficients from data, which greatly reduces the complexity of calculation, and the features are also translation invariant and stable.

\subsection{Train an Encoder With Supervised Contrastive Learning}
The efficacy of the AWSPNet architecture heavily relies on the quality of the feature representations generated by its encoder. To this end, we employ supervised contrastive learning (SCL) for training the encoder module. Our approach offers advantages over traditional supervised learning methods that often rely on direct classification with, for instance, a cross-entropy loss. In general, classification networks trained on specific categories tend to exhibit poor generalization when processing complex and variable radar signals. In contrast, SCL primarily focuses on training robust encoders that are not tailored to a specific task but are instead designed to extract signal features in a generalizable manner, thereby demonstrating superior performance in practical applications. Specifically, SCL explicitly trains the encoder to learn an embedding space where instances from the same class are pulled closer together, while instances from different classes are pushed further apart. This results in more discriminative and robust feature representations that are beneficial for downstream tasks, particularly for the subsequent prototypical network which thrives on well-separated class clusters in the feature space.

Considering recognition accuracy and model complexity, we choose a pre-trained EfficientNetV2 \cite{tanEfficientNetV22021} as the backbone net, integrated through the principles of transfer learning. Utilizing a EfficientNetV2 model pre-trained on large-scale datasets brings several benefits. Firstly, it allows the network to leverage rich, hierarchical features learned from diverse visual data, which can be effectively adapted to the radar signal domain. Secondly, it often accelerates convergence and can reduce the amount of labeled domain-specific data required for effective training, mitigating the risk of overfitting, especially when radar datasets are limited.

Furthermore, the proposed framework exhibits significant flexibility in its encoder design. While EfficientNetV2 serves as a strong baseline, the architecture is modular, permitting the substitution of this backbone with other pre-trained convolutional neural networks. Alternative architectures, such as ShuffleNet\cite{zhangShuffleNet2017}, Resnet50\cite{heDeep2016}, MobileNetV3\cite{howardSearching2019} can be readily integrated depending on specific computational constraints or desired feature characteristics. The comparative performance of these different pre-trained networks within the AWSPNet framework for the current radar target recognition task will be empirically evaluated and discussed in the experimental results section of this paper.

For simplicity, let $g_{\rho} $ represents the nonlinear map of encoder (including the DTCWT), i.e.,
\begin{equation}\label{key}
	g_{\rho}:\mathbf{Z}\in \mathbb{C} ^{Q\times L}\rightarrow \boldsymbol{\alpha }\in \mathbb{R} ^{1024}
\end{equation}
where $\rho  $ represents the learnable parameters in the encoder. Then, the training process via SCL utilizes contrastive loss function\cite{khoslaSupervised2021}. 
\begin{equation}\label{20}
\mathcal{L} =\sum_{i\in I}{\frac{-1}{|P(i)|}\sum_{p\in P(i)}{\log \frac{\exp \left( \boldsymbol{\alpha }_{i}^{T}\boldsymbol{\alpha }_p/\tau \right)}{\sum_{b\in A(i)}{\exp \left( \boldsymbol{\alpha }_{i}^{T}\boldsymbol{\alpha }_b/\tau \right)}}}}
\end{equation}
where $ i\in I\equiv \left\{ 1,\cdots ,3N_B \right\}  $ is the index of an arbitrary augmented sample (In our method, we employ Gaussian noise and random horizontal shift transformations to augment the original data), $ N_B $ is the batch size. $ \tau \in \mathbb{R} ^+  $ is a scalar temperature parameter, and $ A(i)\equiv I\backslash \left\{ i \right\}  $. $ P(i) $ is the set of indices of all positives distinct from $ i $, and $ |P(i)| $ is its cardinality.

\textbf{Remark:} \eqref{20} means that, for a given anchor sample, this loss encourages its feature representation to be closer to other samples from the same class (positive pairs) than to samples from different classes (negative pairs) present in the batch. This is typically achieved by minimizing the distance (e.g., based on cosine similarity) between positive pairs and maximizing it for negative pairs in the learned embedding space. It ensures that the encoder learns highly structured and discriminative features essential for the subsequent few-shot classification performed by the prototypical network.

In the inference stage, the class prediction of a sample $ \mathbf{Z}_t$ (subscript $ t $ represents input matrix changes with time) can be calculated as follows:
\begin{equation}\label{key}
\begin{aligned}
	c_{t}^{*}&=\mathop {\mathrm{arg}\min} \limits_{k\in \left\{ 1,\cdots ,N_c \right\}}\left\| \boldsymbol{\beta }_{t}^{}-\boldsymbol{\alpha }_k \right\| ^2\\
	p_{t}^{*}&=\max_{k\in \left\{ 1,\cdots ,N_c \right\}} \frac{e^{-\left\| \boldsymbol{\beta }_{t}^{}-\boldsymbol{\alpha }_k \right\| ^2}}{\sum_{j=1}^{N_c}{e^{-\left\| \boldsymbol{\beta }_{t}^{}-\boldsymbol{\alpha }_j \right\| ^2}}}\\
\end{aligned}
\end{equation}
where $ \boldsymbol{\beta }_{t}^{}=g_{\rho}\left( \mathbf{Z}_t \right)  $ is the the feature vector of input $ \mathbf{Z}_t$. $ c_{t}^{*} $ is the predicted class label index, and $ p_{t}^{*} $ is the probability of the predicted class $ c_{t}^{*} $. $ N_c $ is the number of class. $ {\alpha }_k, k =1,\cdots,N_c $, is the prototype center for the $ k $-th class, which is expressed as
\begin{equation}\label{key}
	\boldsymbol{\alpha }_k=\frac{1}{\left| S \right|}\sum_{\mathbf{Z}\in S_k}^{}{g_{\rho}\left( \mathbf{Z} \right)}
\end{equation}
where $  S_k$ represents the support set of the $ k $-th class. 

\subsection{Target Recognition and Jamming Suppression Scheme using AWSPNet}
Then, the probability sequence $ p_{t}^{*}, t=1,2,\cdots $ combine with the predicted class label index $ c_{t}^{*} $ can be used to recognize target and suppress jamming. Let $  c_{t}^{*} = 0 $ represents the index of the target label.
Assuming that AWSPNet can recognize $ N_c $ types of interference and that typical point target echoes (which are identical to the transmitted waveform) are regarded as the target class, the resulting probability sequence $ p_{t}^{*} $ with respect to $  c_{t}^{*} = 0 $ forms a fast-time sequence analogous to matched filtering. In this sequence, the peak indicates the target's distance, while interference locations are effectively suppressed. The target recognition method is summarized in algorithm \ref{a1}, and the system framework is shown in Fig.\ref{f1}.
\begin{algorithm}[htbp] 
	\caption{Target Recognition and Jamming Suppression Method using AWSPNet} 
	\label{a1} 
	\begin{algorithmic}[1] 
		\REQUIRE ~~\\ 
		A support set composed of target echo data obtained from simulation or measurement: $\left\{ \mathbf{Z}_i\left| \mathbf{Z}_i\in \mathbb{C} ^{Q\times L},i\in I \right. \right\}   $;\\
		An echo matrix to be detected: $\mathbf{Z}  $;\\
		The well trained prototype network AWSPNet;\\

		\ENSURE ~~\\
		The probability sequence of the predicted target class $p_{t}^{\left( 0 \right)},t=1,2,\cdots ,L-L_0+1$;\\

		\STATE Construct sliding window segmentation of the input echo matrix: $ \mathbf{Z}\left( t \right)  $. 
		
		Let $ \mathbf{Z}=\left[ \mathbf{z}_1,\cdots ,\mathbf{z}_l,\cdots ,\mathbf{z}_L \right] ,\mathbf{z}_l\in \mathbb{C} ^{Q\times 1} $, then $ \mathbf{Z}\left( t \right) =\left[ \mathbf{z}_t,\cdots ,\mathbf{z}_{t+L_0-1} \right] ,t=1,2,\cdots ,L-L_0+1 $ ;
		\STATE Input $ \mathbf{Z}\left( t \right)  $ into AWSPNet to obtain the output probability sequence $  p_{t}^{\left( 0 \right)}$;
		\RETURN $  p_{t}^{\left( 0 \right)}$.
	\end{algorithmic}
\end{algorithm}

\section{Simulation Experiments and Analysis}
\label{s4}
To validate the effectiveness of the proposed method, we conducted comparative and ablation experiments on simulation data generated by the interference signal simulator. All experiments are conducted on an Nvidia RTX 4080Ti GPU with 32 GB of video memory, and implemented in Python + PyTorch. All models in the experiment are optimized with the Adam optimizer.

\subsection{Data Preparation}
\begin{table*}[]
	\caption{Target and Jamming Signal Parameters for Training and Testing Sets}
	\centering
	\begin{tabular}{l|lll}
		\hline
		Signal Type                                                                                             & Parameter                    & Value(training set) & \multicolumn{1}{l|}{Value(testing set)} \\ \hline
		\multirow{7}{*}{\begin{tabular}[c]{@{}l@{}}Target and jamming signal\\ general parameters\end{tabular}} & Carrier frequency            & 10 GHz              & (8$ \sim  $12) GHz                               \\
		& Bandwidth                    & 40 MHz              & (20$ \sim  $60) MHz                             \\
		& Sampling rate                & 48 MHz              & 68 MHz                                  \\
		& Pulse width                  & 1 $ \mu s $                & (1$ \sim  $10) $ \mu s $                                \\
		& PRI                          & 5 $ \mu s $                & (5$ \sim  $50) $ \mu s $                                \\
		& SNR/INR                      & (-6$ \sim  $10) dB          & (-10$ \sim  $15) dB                              \\
		& CNR                          & --                  & (-10$ \sim  $10) dB                              \\ \hline
		Point Target                                                                                            & Number of point              & 1                   & 1$ \sim  $3                                     \\ \hline
		RDFJ, VDFJ, RVDJ                                                                                          & Number of false target: $N_1$ & 3$ \sim  $9                 & 4$ \sim  $10                                    \\ \hline
		CSJ                                                                                                     & Number of comb teeth: $K$     & 10                  & 5$ \sim  $15                                    \\ \hline
		SSJ                                                                                                     & Number of sub-pulse: $N_s$   & 5                   & 4$ \sim  $8                                     \\ \hline
		\multirow{2}{*}{ISFJ, ISRJ}                                                                              & sampling pulse width: $ T_a $         & 0.25 $ \mu s $             & 0.25 $ \mu s $                                 \\
		& sampling period: $ \tau_a $              & 0.05 $ \mu s $             & 0.05 $ \mu s $                                 \\ \hline
		\multirow{2}{*}{RGJ, VGJ, RVGJ}                                                                           & Dragging speed               & 300$ \sim  $600 $ m/s $          & 300$ \sim  $600 $ m/s $                              \\
		& Dragging acceleration        & 50$ \sim  $200 $ m/s^2 $         & 50$ \sim  $200 $ m/s^2 $                               \\ \hline
	\end{tabular}
	\label{table1}
\end{table*}
To comprehensively evaluate the performance of our proposed model, particularly its generalization capability and robustness in complex electromagnetic environments, we constructed distinct training and testing datasets. One is the labeled target and jamming signal fast-slow time raw data for training ($ \mathbf{Z}_i\in \mathbb{C} ^{128\times 241} $, each with 1000 samples), the other is test dataset consists of multi-target and (compound) jamming signal under different parameter (carrier frequency, velocity, range, sampling, etc). The specific parameter configurations for these datasets are detailed in Table \ref{table1}.

The training dataset was generated using fixed values or narrow parameter ranges (e.g., a fixed carrier frequency of 10 GHz and a SNR range of -6 dB to 10 dB). This approach is designed to provide the model with a baseline learning environment.

In contrast, the testing dataset features significantly broadened parameter ranges (e.g., a carrier frequency from 8 GHz to 12 GHz and an SNR/Interference-to-Noise Ratio (INR) extending from -10 dB to 15 dB) and incorporates various typical interference patterns. It assesses the model's true performance when confronted with more challenging and previously unseen signal conditions. As confirmed by subsequent experimental results, our model demonstrates superior performance compared to baseline methods under this challenging condition.

\subsection{Comparative Experiment and Ablation Experiments}
In this section, the average recognition accuracy, F1 score (the purpose of the F1 score is to combine precision and recall to evaluate the overall performance of a model) and average time cost of inference are chosen as the metrics to evaluate the performance. The higher the average recognition accuracy and F1 score, the better the performance. All experimental results are shown in Table \ref{table2}. The default experimental parameters are: batch size = 64, epochs = 50, learning rate = 0.0005.

\begin{table*}[]
	\caption{Results of Comparative Experiments and Ablation Experiments}
	\centering
	\begin{tabular}{l|l|llll|llll|l|l}
		\hline
		\multirow{2}{*}{\begin{tabular}[c]{@{}l@{}}\textbf{Experiment} \\ \textbf{purpose}\end{tabular}}                                                                                                                     & \multirow{2}{*}{\textbf{Method}}                                                                                                    & \multicolumn{4}{l|}{\textbf{Accuracy} under different SNR(\%)}             & \multicolumn{4}{l|}{\textbf{F1 Score} under different SNR}             & \multirow{2}{*}{\textbf{Paras} (M)} & \multirow{2}{*}{\begin{tabular}[c]{@{}l@{}}\textbf{Inference} \\ \textbf{Time} (s)\end{tabular}} \\ \cline{3-10}
		&                                                                                                                            & -6 dB          & -3 dB          & 0 dB           & 10 dB          & -6 dB         & -3 dB         & 0 dB          & 10 dB         &                            &                                                                               \\ \hline \hline 
		\multirow{5}{*}{\begin{tabular}[c]{@{}l@{}}Method \\ comparison\\ (trained with \\ SNR=10 dB \\ data sets,\\ support with \\ the same)\end{tabular}}    & Prior-K-NN                                                                                                                 & 13.76          & 20.16          & 30.17          & 99.11          & 0.06          & 0.10          & 0.21          & 0.99          & 20                         & 0.18                                                                          \\
		& MS-WSN                                                                                                                     & 18.06          & 19.15          & 22.97          & 99.21          & 0.07          & 0.07          & 0.23          & 0.99          & 5.4                        & 0.05                                                                          \\
		& PSPNet                                                                                                                     & 38.55          & 40.55          & 57.31          & 99.21          & 0.39          & 0.40          & 0.56          & 0.99          & 0.202                      & 0.0006                                                                        \\ \cline{2-12} 
		& \begin{tabular}[c]{@{}l@{}}AWSPNet(proposed)\\ with EfficientV2\end{tabular}                                               & 52.44          & 75.25          & 93.56          & 99.91          & 0.46          & 0.75          & 0.93          & 0.99          & 24                         & 0.19                                                                          \\ \cline{2-2}
		& \begin{tabular}[c]{@{}l@{}}AWSPNet\\ with Resnet50\end{tabular}                                                            & 45.32          & 72.54          & 91.91          & 99.91          & 0.45          & 0.72          & 0.91          & 0.99          & 25.5                       & 0.20                                                                          \\ \hline\hline
		\multirow{3}{*}{\begin{tabular}[c]{@{}l@{}}Ablation \\ Experiments\\ (trained with \\ SNR=10 dB \\ data sets,\\ support with \\ the same)\end{tabular}} & \begin{tabular}[c]{@{}l@{}}AWSPNet \\ without \\ DTCWT module\end{tabular}                                                 & 42.12          & 72.31          & 92.01          & 99.01          & 0.42          & 0.71          & 0.92          & 0.99          & 24                         & 0.16                                                                          \\ \cline{2-2}
		& \begin{tabular}[c]{@{}l@{}}AWSPNet \\ without \\ Attention module\end{tabular}                                             & 42.18          & 72.14          & 93.06          & 99.91          & 0.37          & 0.72          & 0.93          & 0.99          & --                         & 0.19                                                                          \\ \cline{2-2}
		& \begin{tabular}[c]{@{}l@{}}AWSPNet \\ without \\ pre-train weights\end{tabular}                                            & 15.95          & 24.13          & 37             & 93.45          & 0.06          & 0.16          & 0.32          & 0.93          & --                         & 0.20                                                                          \\ \hline\hline
		\multirow{3}{*}{\begin{tabular}[c]{@{}l@{}}Comparison of \\ training and \\ support set \\ distribution\end{tabular}}                                   & \begin{tabular}[c]{@{}l@{}}AWSPNet (trained \\ with multi-SNR \\ mixed data sets)\end{tabular}                             & \textbf{90.45} & \textbf{94.33} & \textbf{95.78} & \textbf{95.01} & \textbf{0.90} & \textbf{0.94} & \textbf{0.95} & \textbf{0.95} & 24                         & 0.20                                                                          \\ \cline{2-2}
		& \begin{tabular}[c]{@{}l@{}}AWSPNet (trained \\ with SNR=-6 dB \\ data sets)\end{tabular}                                   & 82.42          & 61.91          & 55.95          & 50.11          & 0.81          & 0.63          & 0.51          & 0.5           & --                         & 0.20                                                                          \\ \cline{2-2}
		& \begin{tabular}[c]{@{}l@{}}AWSPNet (trained \\ with SNR=10 dB \\ data sets, support with \\ multi-SNR data)\end{tabular} & 42.44          & 55.15          & 91.49          & 99.11          & 0.37          & 0.50          & 0.91          & 0.99          & --                         & 0.20                                                                          \\ \hline
	\end{tabular}
\label{table2}
\end{table*}

\subsubsection{Comparative experiments}
The primary objective of this experiment is to evaluate the performance of our proposed AWSPNet method under various signal-to-noise ratio (SNR) conditions and to benchmark it against several baseline methods, including the prior-knowledge-guided neural network based on supervised contrastive learning (Prior-K-NN) method\cite{liuPriorKnowledgeGuided2024}, multiple-scale wavelet scattering network (MS-WSN) method \cite{wangApplication2024}, pre-training and self-supervised fine-tuning-based prototypical network (PSPNet) method \cite{xiaoPSPNet2024}. 

As detailed in the first five rows of Table \ref{table2}, all methods demonstrate nearly flawless performance under the ideal condition of 10 dB SNR, achieving accuracy and F1-scores of approximately 0.99. This suggests that all tested models can effectively complete the task when noise level is minimal. However, as the SNR progressively decreases from 10 dB to -6 dB, a significant disparity in performance emerges, highlighting the varying robustness of the models in high-noise environments.

Specifically, the three baseline methods exhibit severe performance degradation at low SNRs. The Prior-K-NN and MS-WSN methods are rendered almost ineffective at -3 dB and -6 dB. At an SNR of -6 dB, their accuracies fall to a mere 13.76\% and 18.06\%, with corresponding F1-scores of only 0.06 and 0.07, respectively. This demonstrates that traditional methods and early-generation network models struggle to counteract strong noise interference. While PSPNet outperforms the other two baselines, achieving a 38.55\% accuracy at -6 dB, it still suffers a drastic performance decline. Notably, PSPNet's advantages lie in its exceptionally low parameter count (0.202 M) and rapid inference speed (0.0006 s), indicating its efficiency as a lightweight model. This efficiency, however, is achieved at the expense of robustness against noise.

In contrast, our proposed AWSPNet, utilizing either an EfficientV2 or a Resnet50 backbone, demonstrates superior performance and robustness in low-SNR conditions. At -6 dB, the AWSPNet model with EfficientV2 achieves an accuracy of 52.44\% and an F1-score of 0.46, while the Resnet50-based variant reaches 45.32\% accuracy and a 0.45 F1-score. These results are significantly higher than those of all baseline methods. As the SNR increases to 0 dB, the superiority of AWSPNet becomes even more pronounced, with accuracies reaching 93.56\% (EfficientV2) and 91.91\% (Resnet50).

In summary, utilizing different pre-trained backbones for our AWSPNet architecture yields robust recognition performance. When considering the trade-off between recognition accuracy and model complexity, the EfficientV2 version of AWSPNet offers a more advantageous balance. However, the proposed method exhibits suboptimal performance under low SNR conditions. This performance degradation is a direct consequence of the training dataset's composition, which exclusively features data at a single SNR level of 10 dB. The use of a single SNR value results in a non-uniform data distribution, thereby limiting the model's robustness and generalization capabilities in high-noise environments.
\subsubsection{Ablation experiments}
To validate the effectiveness and necessity of the key components within our proposed AWSPNet model, we conducted a series of ablation studies. The results are shown in the sixth to eighth rows of Table \ref{table2}. We systematically evaluated the change in model performance by individually removing its core modules: the DTCWT module, the Attention module, and the pre-trained weights. The results, when compared to the performance of the complete AWSPNet model (the fourth to eighth rows of Table \ref{table2}), clearly reveal the significant contribution of each component.

1). Efficacy of the DTCWT module: 
Upon removing the DTCWT module, the model's performance experienced a significant decline. Specifically, under the -6 dB SNR condition, the accuracy dropped from 52.44\% to 42.12\%, and the F1-score decreased from 0.46 to 0.42. Similar performance degradation was observed at -3 dB and 0 dB SNR. This result illustrates the critical role of the DTCWT module, which effectively extracts robust time-frequency domain features from the signal under background noise. It provides the network with more informative and resilient input, thereby significantly enhancing the model's recognition capability in low-SNR environments. Although its removal reduces the inference time from 0.19s to 0.16s, the substantial performance improvement it provides justifies this computational overhead.

2). Efficacy of the attention module: 
When the attention module was removed, the model's performance was also severely impacted, with a magnitude of degradation comparable to that of removing the DTCWT module. At a -6 dB SNR, the accuracy decreased to 42.18\%. This indicates that the attention mechanism plays a vital role in enabling the model to focus on critical signal features while simultaneously suppressing irrelevant noise. Under high-noise condition, the attention module guides the allocation of computational resources to the most information-rich regions, which is indispensable for achieving precise classification and ensuring model robustness.

3). Efficacy of pre-trained weights: 
When the pre-trained weights were removed and the model was trained from random initialization, its performance suffered a precipitous decline. At -6 dB SNR, the accuracy plummeted to 15.95\% with an F1-score of just 0.06—a result nearly equivalent to random guessing. Even under the more favorable 0 dB SNR condition, the accuracy reached only 37\%, far below the full model's 93.56\%. This outcome clearly demonstrates that for complex deep learning models, transfer learning (i.e., using pre-trained weights) is of paramount importance. The pre-trained weights provide an excellent initialization point, allowing the model to leverage general features learned on large-scale datasets. This approach helps prevent the model from getting trapped in local optima and dramatically improves both its convergence speed and final performance.

In summary, the ablation studies confirm that the DTCWT module, the attention module, and the pre-trained weights collectively form the cornerstone of AWSPNet's high performance. The removal of any of these components leads to a significant degradation in model performance, particularly in low-SNR environments. The DTCWT and attention modules are complementary: the DTCWT module extracts robust features from the source signal, while the attention module subsequently refines and focuses on these features, jointly ensuring the model's powerful noise immunity. Finally, the adoption of a transfer learning strategy via pre-trained weights is a critical factor in achieving high performance, especially at low SNRs.

\subsection{Visual analysis of the output of AWSPNet Under different SNR Condition}
\begin{figure*}[htbp]
	\centering
	\subfigure[]{
			\includegraphics[width=0.30\textwidth]{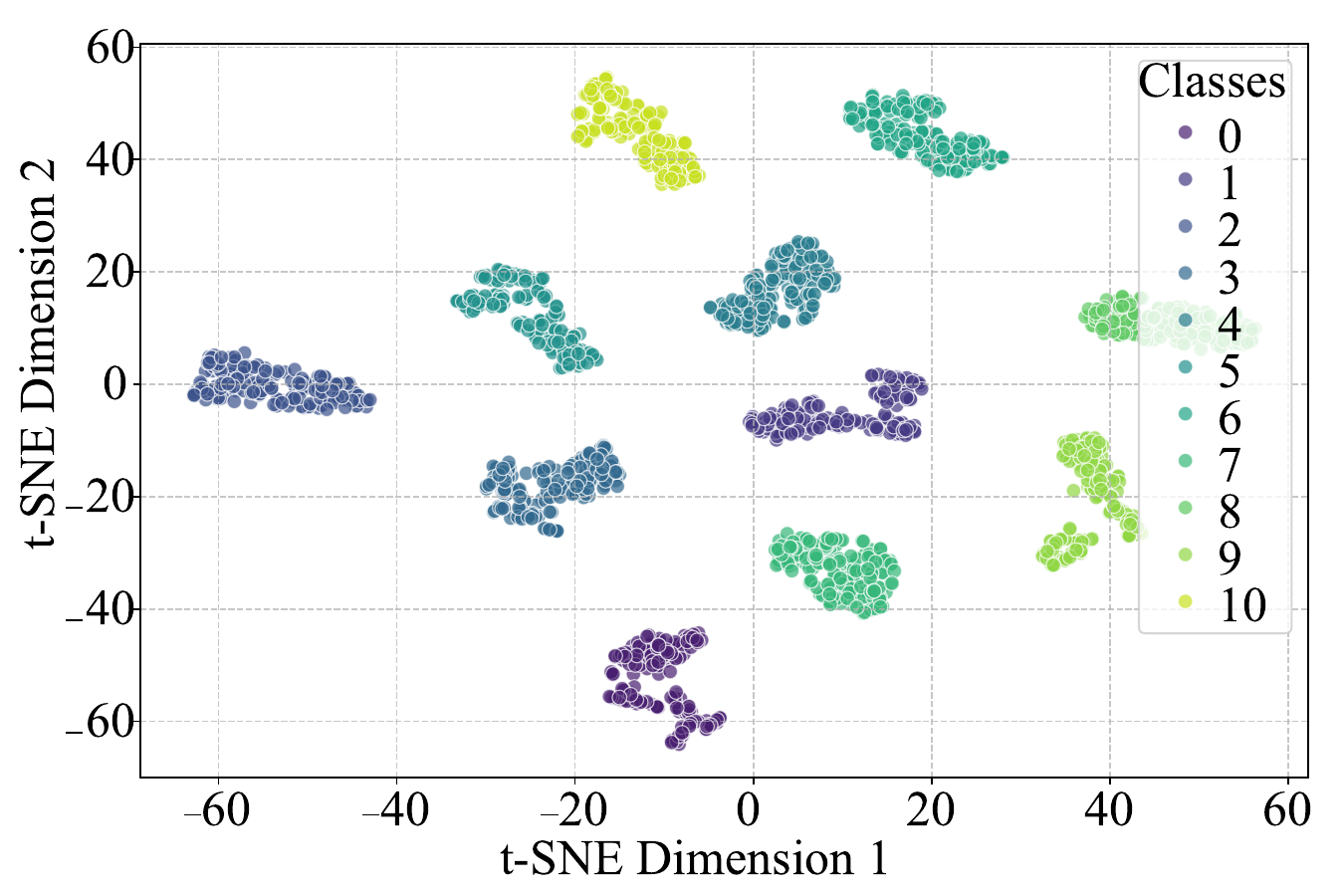}\label{f6a}
		}
	\,
	\subfigure[]{
			\includegraphics[width=0.30\textwidth]{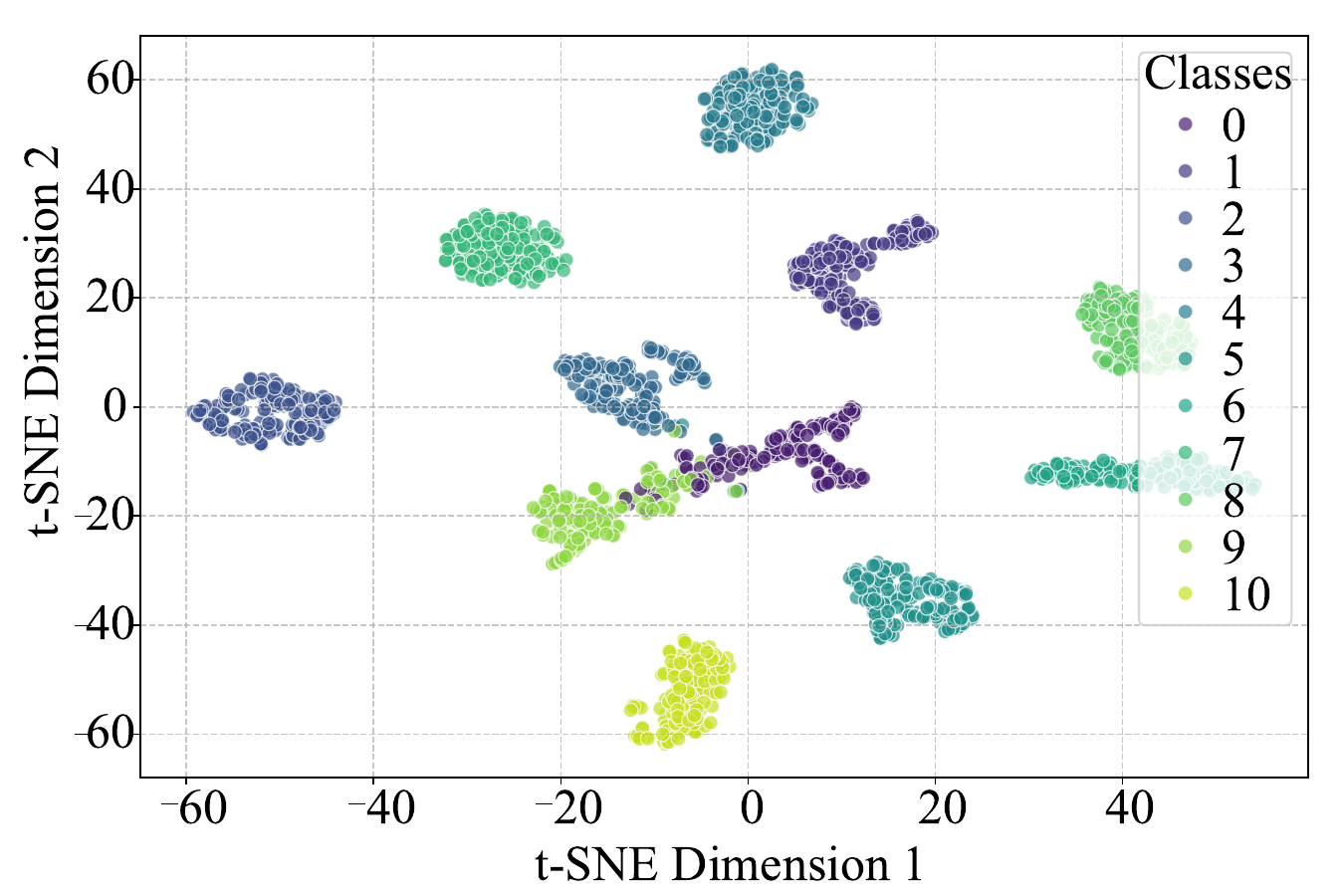}\label{f6b}
		}
	\,
	\subfigure[]{
			\includegraphics[width=0.30\textwidth]{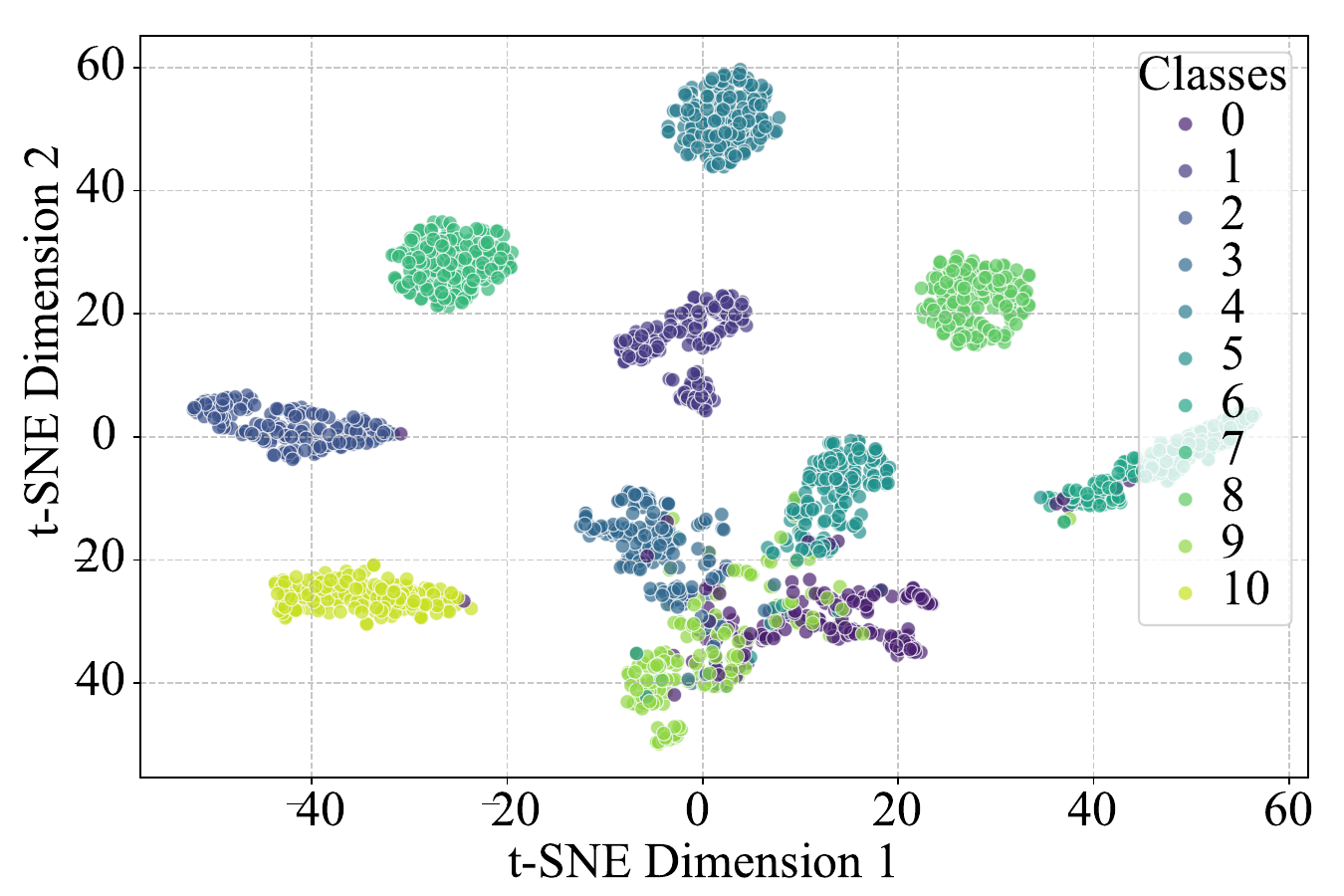}\label{f6c}
		}
	\,
		\subfigure[]{
			\includegraphics[width=0.30\textwidth]{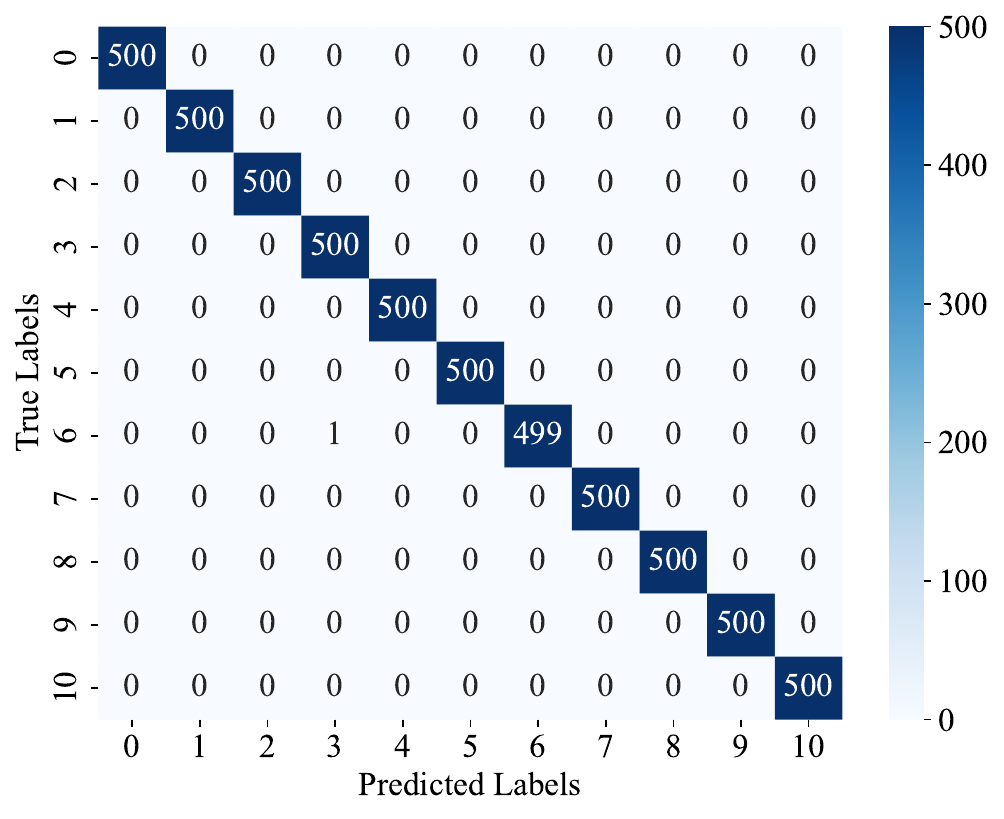}\label{f6d}
		}
	\,
	\subfigure[]{
			\includegraphics[width=0.30\textwidth]{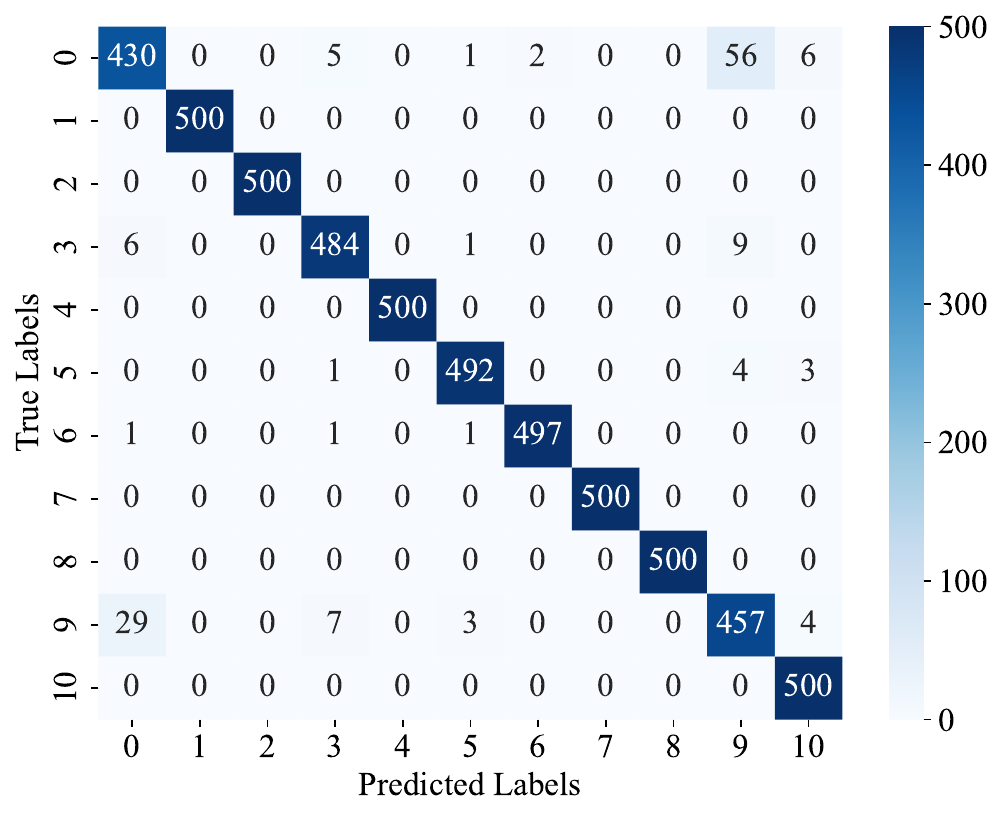}\label{f6e}
		}
	\,
	\subfigure[]{
			\includegraphics[width=0.30\textwidth]{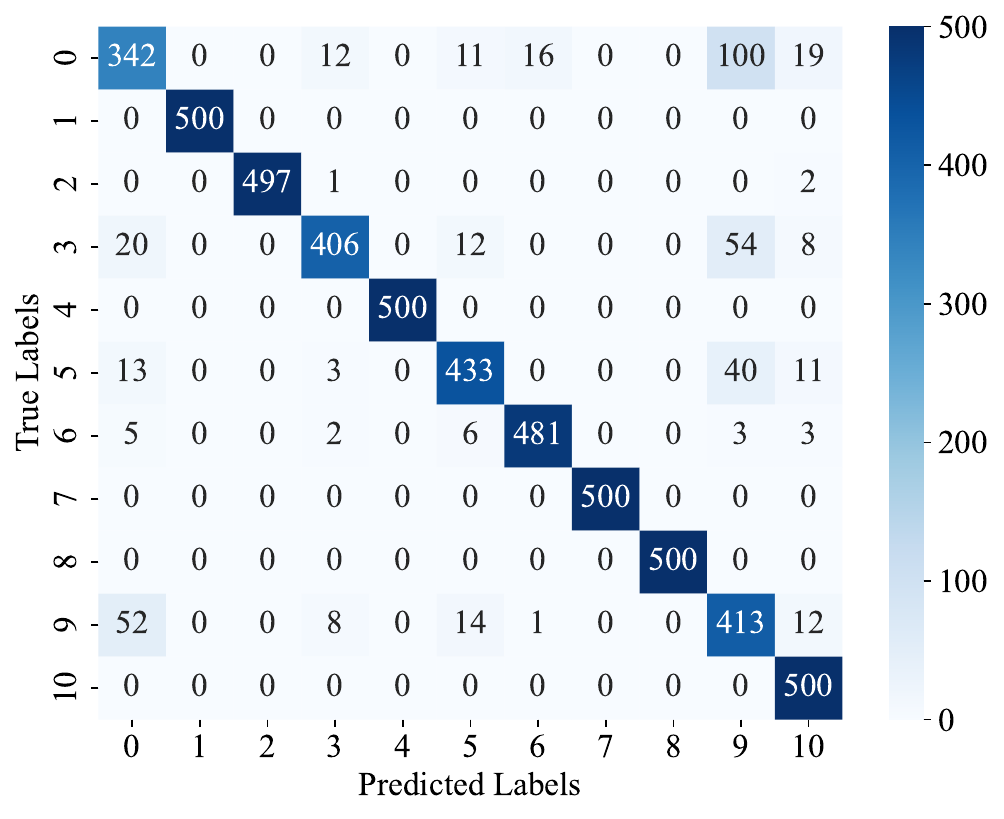}\label{f6f}
		}
	\caption{Model performance (using EfficientV2 as pre-train backbone net) on the test dataset at various SNRs. It presents the confusion matrices for the classification results and visualizations of the corresponding output distributions using the t-SNE algorithm. The eleven class labels, indexed from 0 to 10, are defined as follows: point target, RDFJ, VDFJ, RVDJ, ISFJ, ISRJ, CSJ, RGJ, VGJ, RVGJ, and SSJ. The t-SNE visualizations illustrate the separability of the predicted classes at SNRs of (a) 10 dB, (b) -3 dB, and (c) -10 dB. The corresponding confusion matrices detail the classification performance at these same SNR levels: (d) 10 dB, (e) -3 dB, and (f) -10 dB.}
	\label{f6}
\end{figure*}
To provide an intuitive assessment of the model's classification performance under varying SNRs, in this section we employed the t-SNE algorithm \cite{vandermaaten08a} to visualize the dimensionality-reduced output features of the test set. This analysis is complemented by corresponding confusion matrices, as shown in Fig.\ref{f6}. Fig.\ref{f6a}, \ref{f6b}, and \ref{f6c} illustrate the t-SNE feature distributions at SNRs of 10 dB, -3 dB, and -10 dB, respectively, while Fig.\ref{f6d}, \ref{f6e}, and \ref{f6f} present the corresponding confusion matrices for these conditions.

At a high SNR of 10 dB, as shown in Fig.\ref{f6a}, the features extracted by the model exhibit exceptional separability. The feature points for each signal class form well-defined, compact, and distinct clusters in the two-dimensional space, with significant distance between different classes. This ideal feature distribution directly corresponds to the near-perfect classification accuracy shown in Fig.\ref{f6d}. The confusion matrix displays a strong diagonal structure, with the values on the diagonal (500 samples per class) indicating that nearly all samples were correctly classified, thus proving the model's outstanding performance in low-noise environments.

When the SNR decreases to -3 dB, the model's performance begins to be challenged by the increasing noise. The t-SNE visualization in Fig.\ref{f6b} shows that while most classes remain well-separated, the intra-class distributions have become more diffuse. Notably, the feature cluster boundaries for Class 0 (point target) and Class 9 (RVGJ) exhibit partial overlap, indicating the model's difficulty in completely distinguishing these two signals at this noise level. This phenomenon is quantitatively validated by the confusion matrix in Fig.\ref{f6e}, which shows that 56 samples from Class 0 were misclassified as Class 9, and 29 samples from Class 9 were misclassified as Class 0, directly contributing to classification inaccuracy.

In the high-noise environment of -10 dB SNR, the t-SNE plot in Fig.\ref{f6c} reveals a significant increase in the dispersion of feature clusters and an expansion of the overlap regions between different classes. The classification boundaries for Class 0 (point target), Class 3 (ISFJ), Class 5 (VJG), and Class 9 (RVGJ) become particularly blurred with substantial overlap. This confusion in the feature space leads to the significant performance degradation documented in Fig.\ref{f6f}. The confusion matrix shows a sharp increase in misclassifications between these classes; for example, 100 samples from Class 0 were incorrectly predicted as Class 9, while 52 samples from Class 9 were mistaken for Class 0, with similar errors occurring among the other overlapping classes.

In summary, the t-SNE visualizations and confusion matrices are in strong agreement, collectively revealing a clear pattern of model performance relative to SNR:
At high SNRs, the learned features are highly separable, enabling precise classification.
As noise intensifies, the features of different signals become progressively indistinct, reducing the inter-class distance in the feature space and leading to the cluster overlap observed in the t-SNE plots.
This feature space overlap is the direct cause of classification errors. The specific misclassifications between classes in the confusion matrices directly correspond to the regions of blurred boundaries in the t-SNE visualizations.
Finally, despite the performance decline under the harsh -10 dB condition, the model retains a commendable recognition capability for the majority of classes, demonstrating its strong overall robustness.
\subsection{Visual analysis of the output of each module and pre-train backbone net of AWSPNet}
\begin{figure*}[htbp]
	\centering
	\subfigure[]{
		\includegraphics[width=0.30\textwidth]{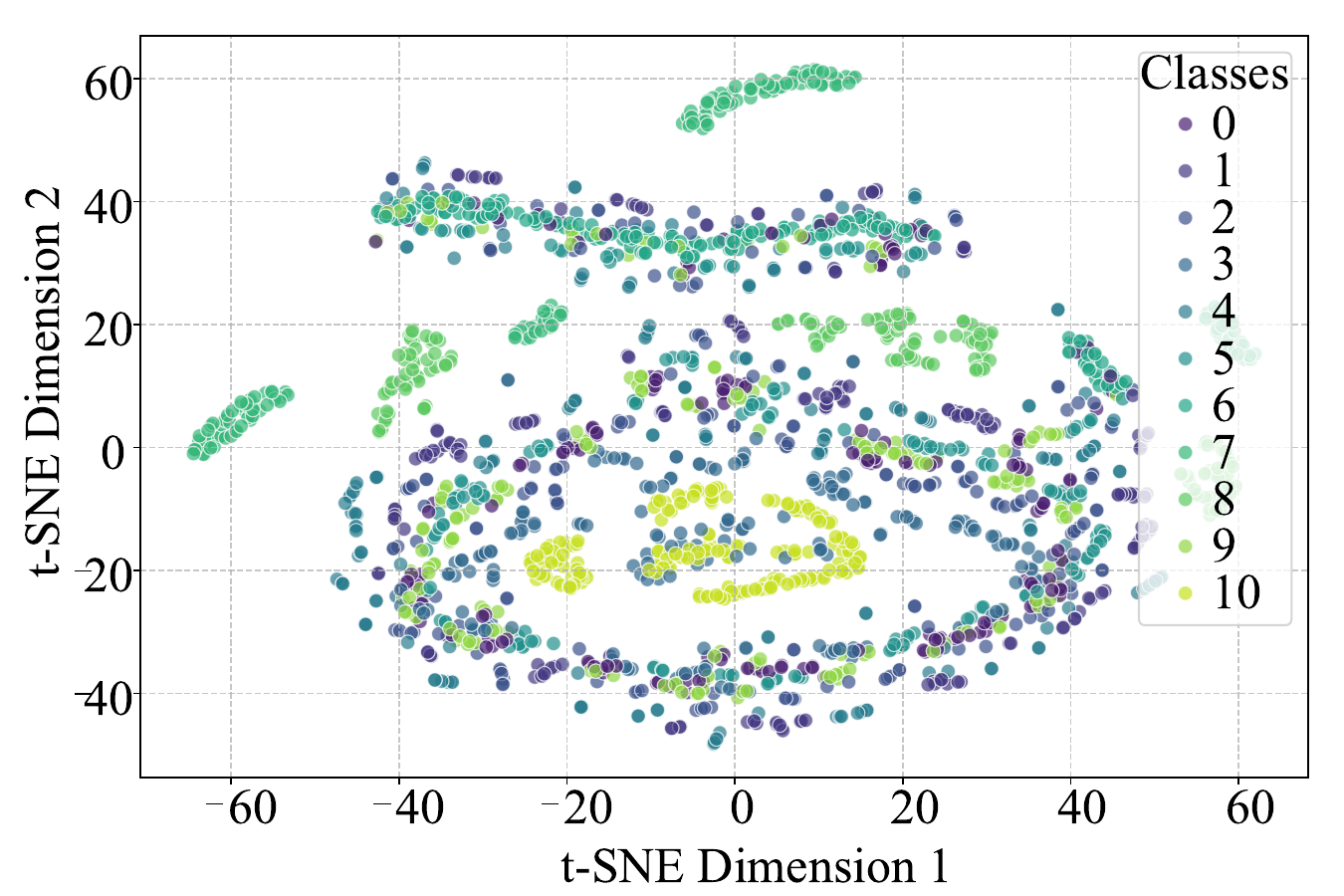}\label{f7a}
	}
	\,
	\subfigure[]{
		\includegraphics[width=0.30\textwidth]{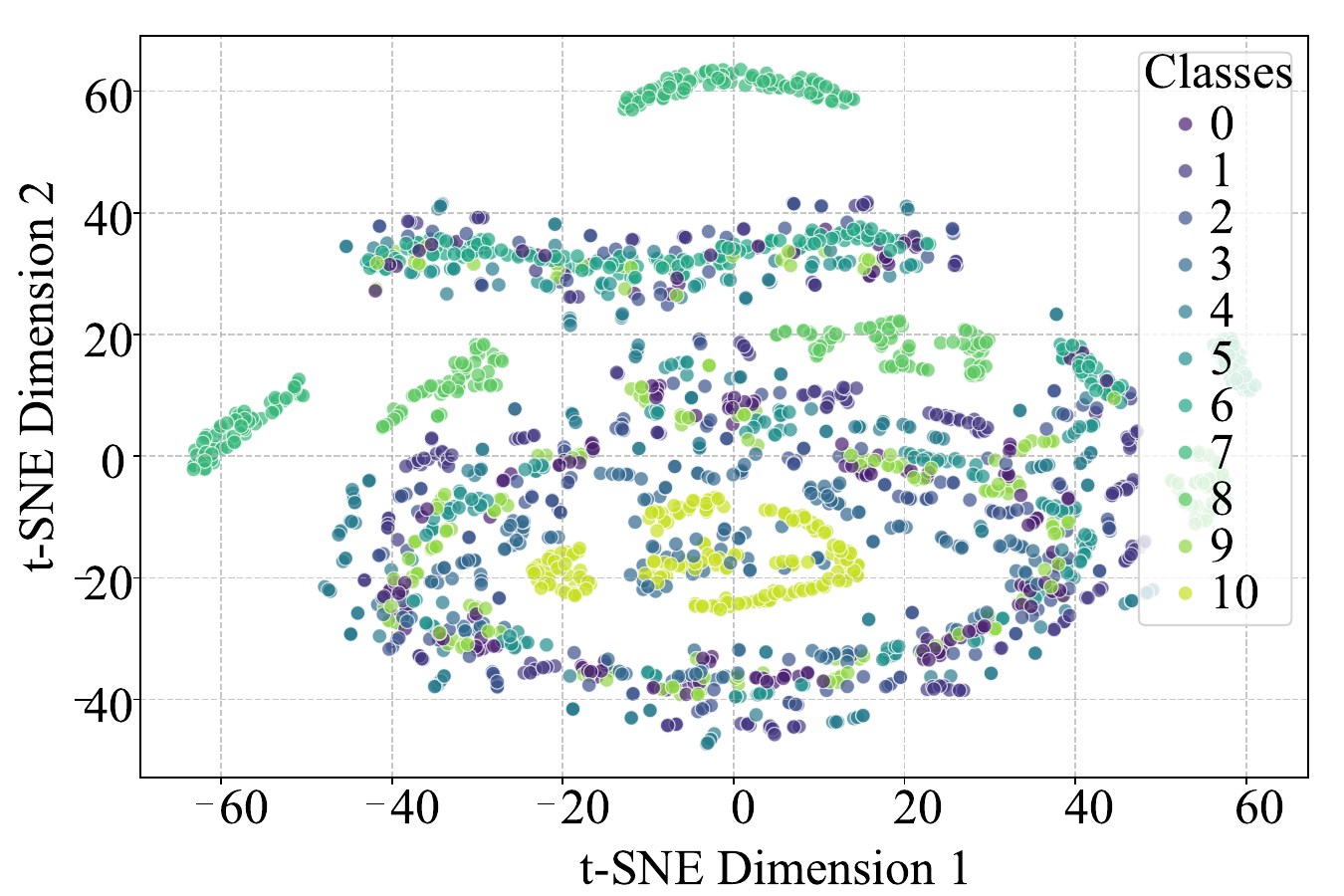}\label{f7b}
	}
	\,
		\subfigure[]{
		\includegraphics[width=0.30\textwidth]{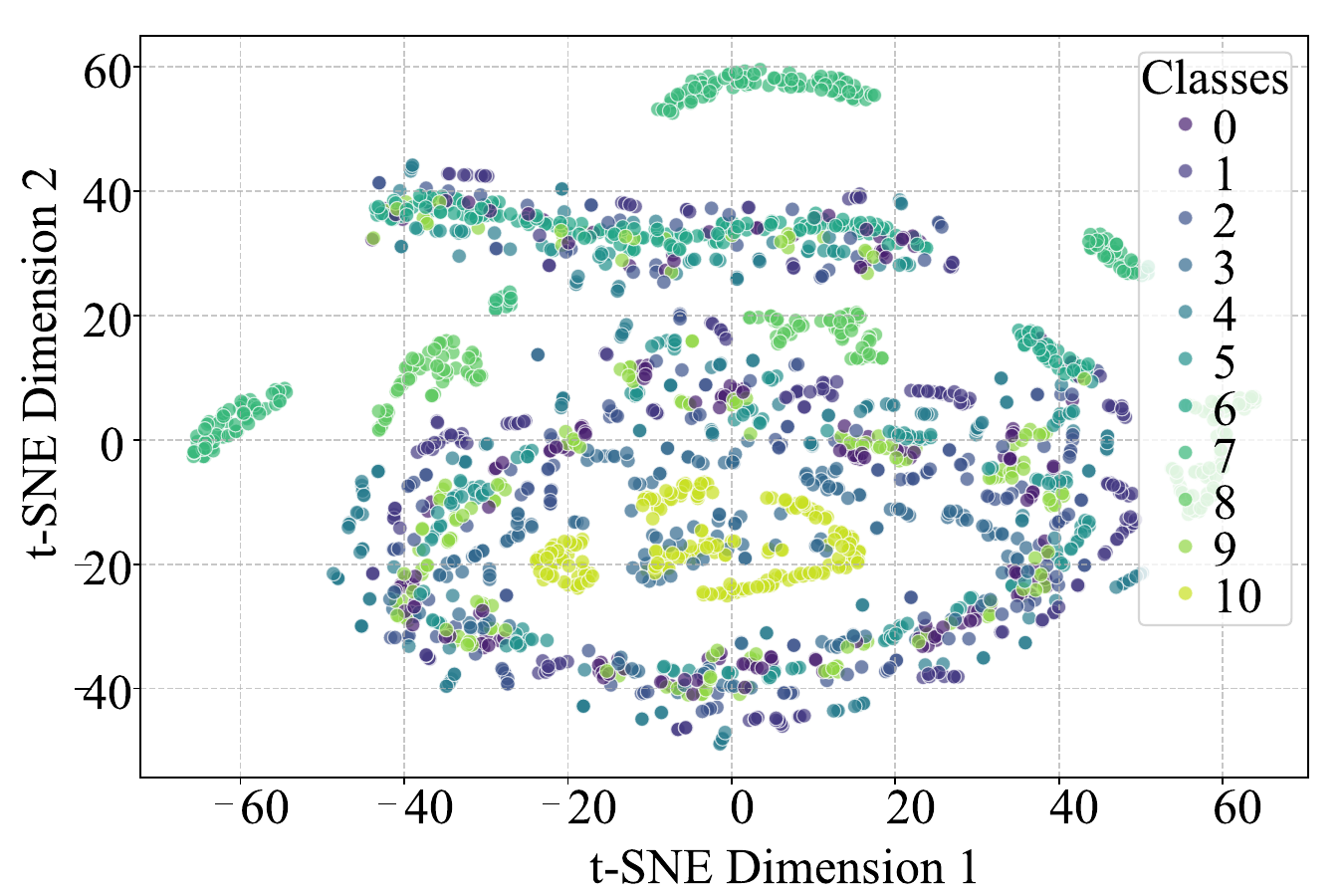}\label{f7c}
	}
\,
	\subfigure[]{
		\includegraphics[width=0.30\textwidth]{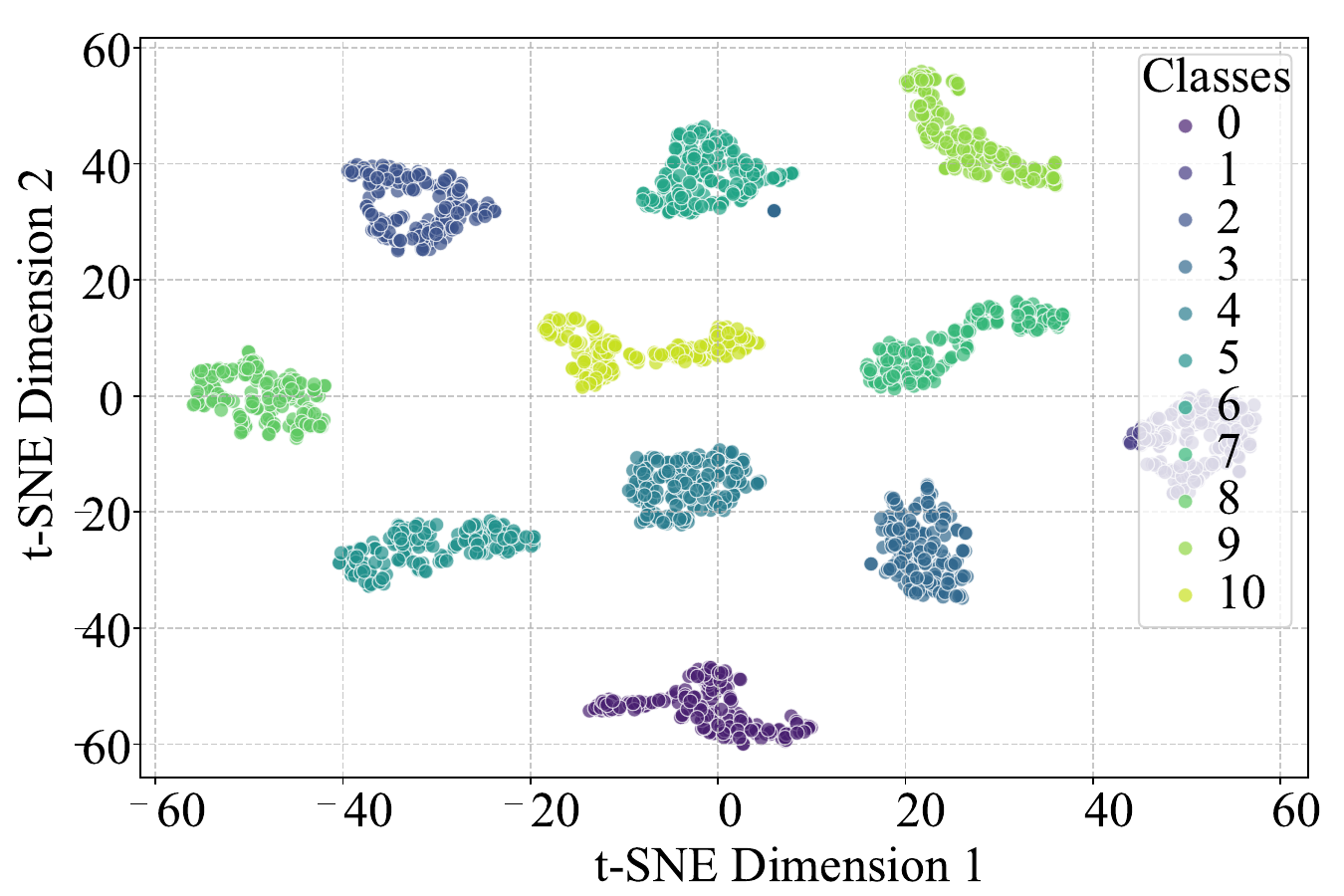}\label{f7d}
	}
	\,
	\subfigure[]{
		\includegraphics[width=0.30\textwidth]{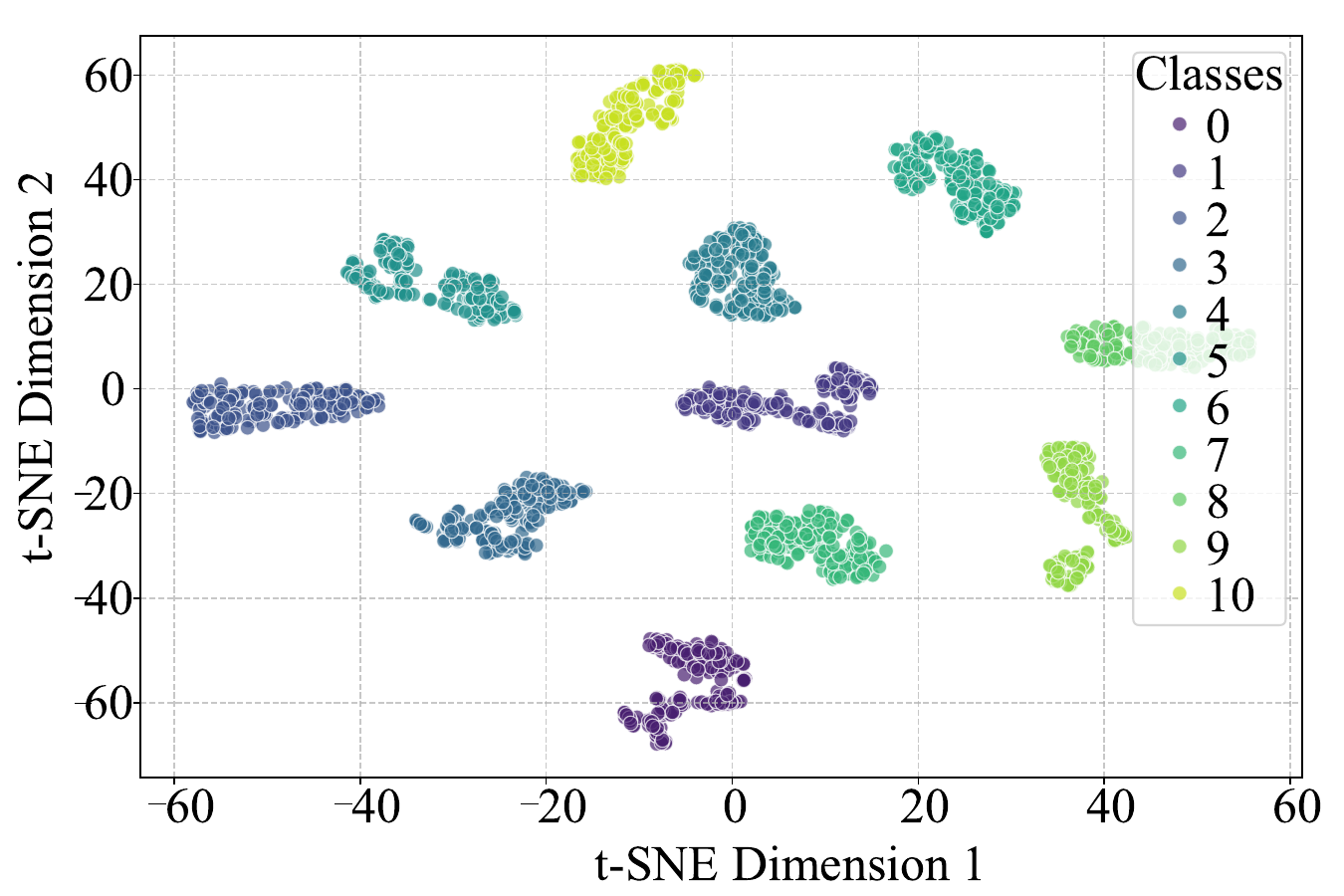}\label{f7e}
	}
\,
	\subfigure[]{
	\includegraphics[width=0.30\textwidth]{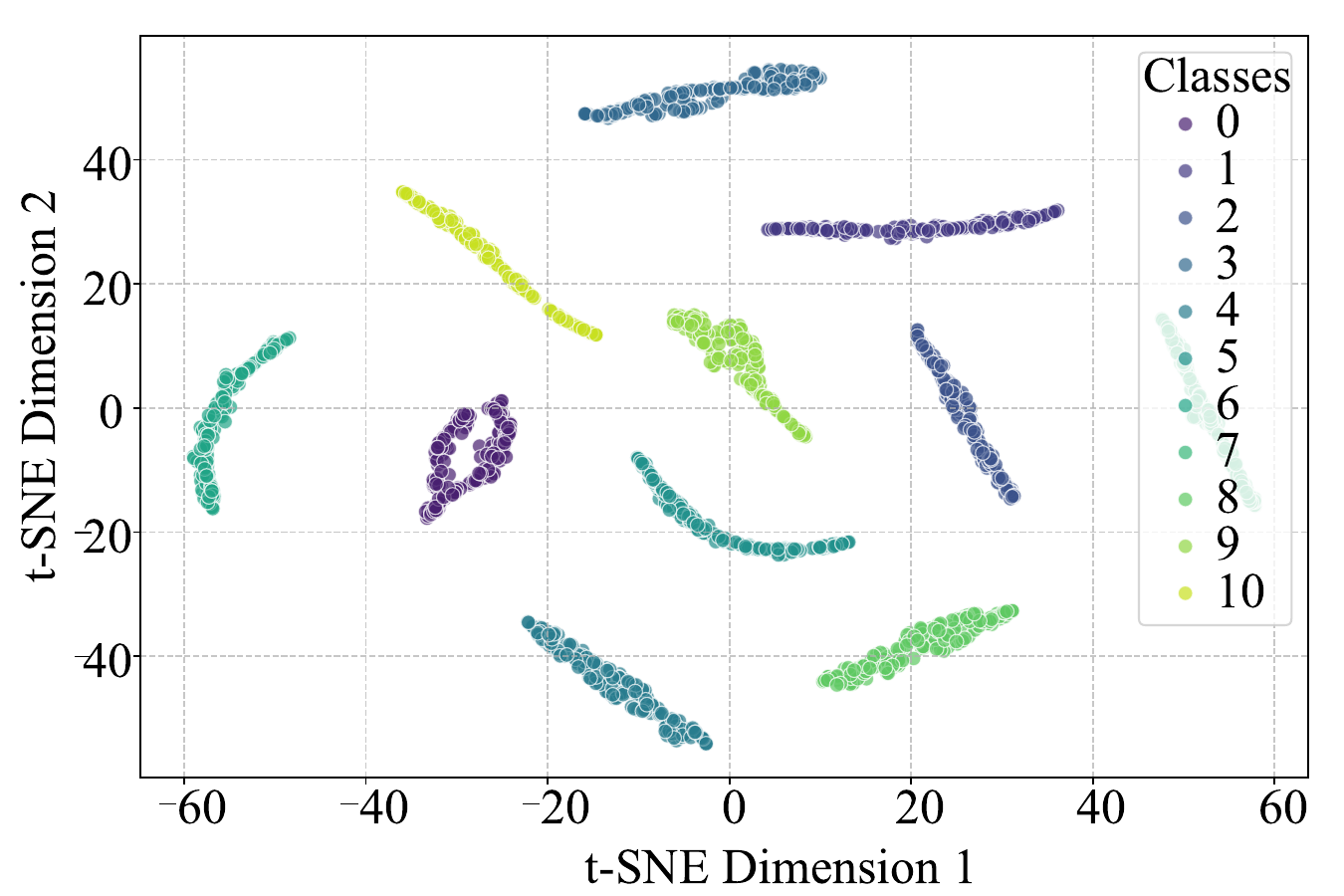}\label{f7f}
}
\,
\subfigure[]{
	\includegraphics[width=0.30\textwidth]{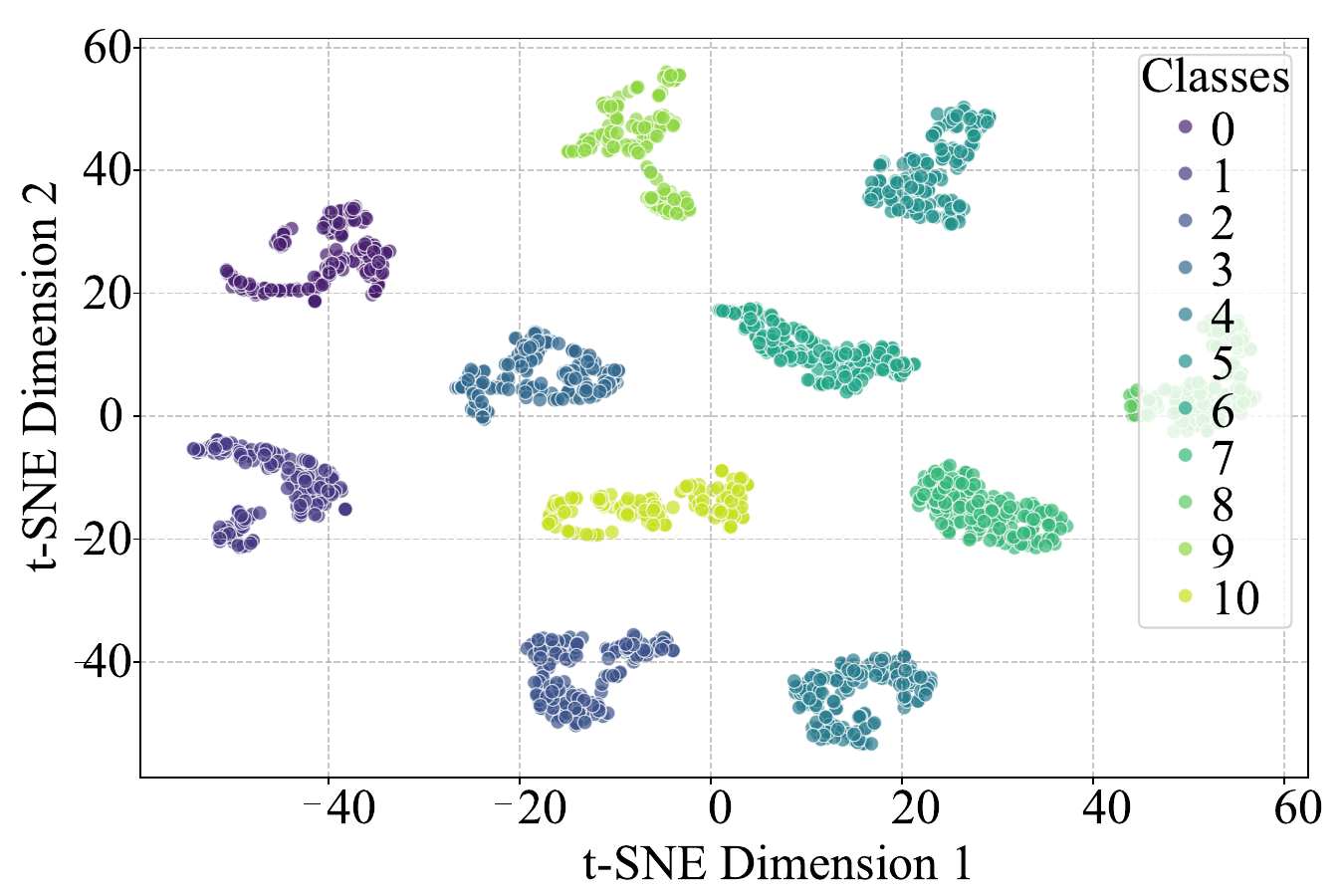}\label{f7g}
}
\,
\subfigure[]{
	\includegraphics[width=0.30\textwidth]{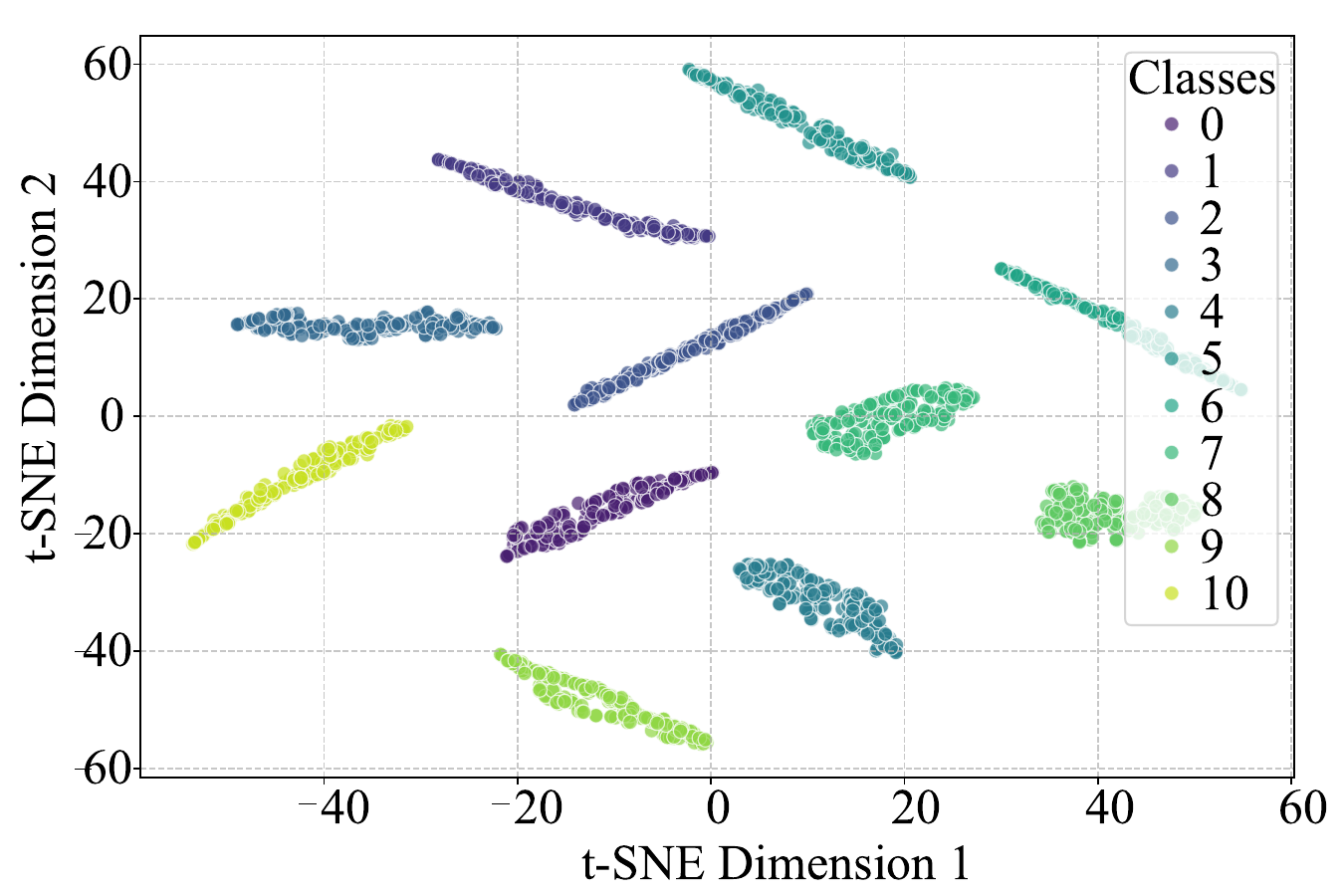}\label{f7h}
}
\,
\subfigure[]{
	\includegraphics[width=0.30\textwidth]{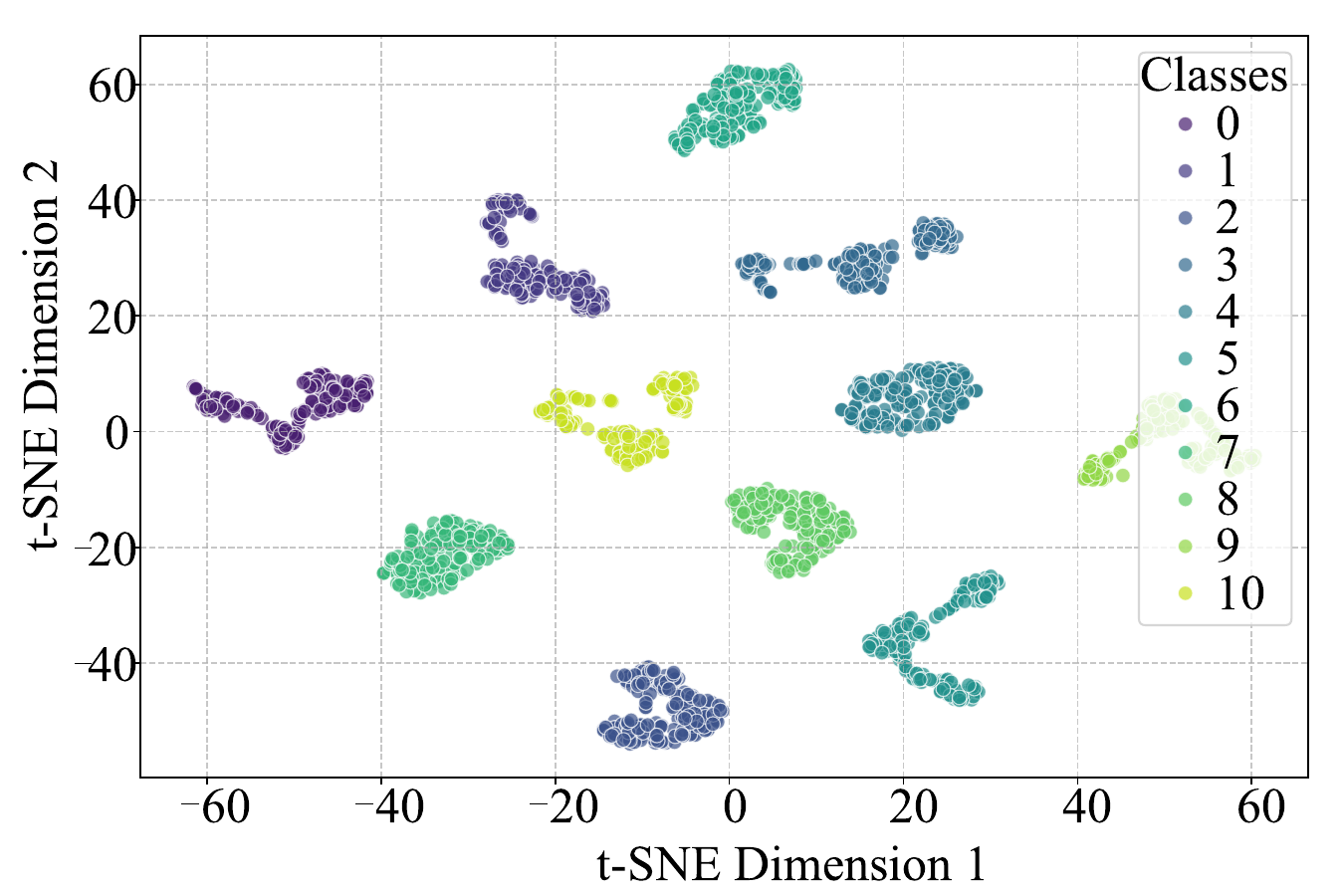}\label{f7i}
}
	\caption{ t-SNE visualization of feature distributions at different stages of the AWSPNet model with various backbone networks, which illustrates the evolution of feature separability within the AWSPNet architecture, visualized using the t-SNE algorithm on test data with a 10 dB SNR. The eleven class labels, indexed from 0 to 10, are defined as: Point target, RDFJ, VDFJ, RVDJ, ISFJ, ISRJ, CSJ, RGJ, VGJ, RVGJ, and SSJ. The sub-figures show the feature distributions after specific modules: (a) After the DTCWT module. As DTCWT is a non-learnable layer with fixed parameters, this output is identical regardless of the subsequent backbone network. (b-c) After the Attention module, using (b) EfficientNetV2\cite{tanEfficientNetV22021} and (c) ResNet50\cite{heDeep2016} as the backbone.
		(d-i) After the final Finetuning net module, showcasing the final classification separability with different pre-train backbones: (d) ConvNeXtV2\cite{wooConvNeXtV2Codesigning2023}, (e) EfficientNetV2\cite{tanEfficientNetV22021} , (f) ResNet50\cite{heDeep2016}, (g) ShuffleNet\cite{zhangShuffleNet2017}, (h) ResNeXt\cite{xieAggregatedResidualTransformations2017}, and (i) MobileNetV3\cite{howardSearching2019}. }
	\label{f7}
\end{figure*}
To conduct an in-depth investigation of feature separability at different stages within the AWSPNet model and to assess the influence of various pre-trained backbones, we visualized the feature distributions from the 10 dB SNR test data using the t-SNE algorithm, as shown in Fig.\ref{f7}.

Fig.\ref{f7a} displays the feature distribution after the raw signals have been processed only by the DTCWT module. It is evident from the plot that the resulting features do not effectively discriminate between the different signal classes. The feature clusters for most classes present as diffuse, point-cloud-like distributions that are both widespread and severely overlapping. This visually demonstrates that these signals possess a high degree of similarity in the wavelet domain, rendering them difficult to classify using traditional signal processing methods alone.

Subsequently, Fig.\ref{f7b} shows the result after these DTCWT features are passed through an Attention module (using a pre-trained EfficientV2 backbone). A comparison between Fig.\ref{f7a} and Fig.\ref{f7b} reveals only a marginal improvement in the feature space distribution, suggesting that the attention module by itself is insufficient to significantly enhance class separability. Furthermore, the minimal difference between Fig.\ref{f7b} and Fig.\ref{f7c} (using a pre-trained ResNet50 backbone) indicates that although the backbone architectures differ, the attention mechanisms they train learn to focus on similar salient regions of the signal.

In contrast, feature separability is well enhanced after the fine-tuning stage with a pre-trained backbone. The comparison between Fig.\ref{f7a}, \ref{f7b}, \ref{f7c} and Fig.\ref{f7d}-Fig. \ref{f7i} clearly shows that the pre-trained module plays a decisive role in constructing a highly separable feature space, which is critical to the success of the entire AWSPNet model. A deeper inspection of the final feature distributions in Fig. \ref{f7d} through Fig.\ref{f7i} reveals a notable pattern: the morphology of the feature clusters can be categorized into two distinct types, and these types correlate with the model's robustness to noise.

Linear Distributions: represented by Fig.\ref{f7f} (ResNet50) and Fig.\ref{f7h} (ResNeXt\cite{xieAggregatedResidualTransformations2017}), where the feature clusters for each class manifest as elongated, strip-like linear patterns. Clustered distributions: represented by Fig.\ref{f7e} (EfficientNetV2\cite{tanEfficientNetV22021}), Fig.\ref{f7d} (ConvNeXtV2\cite{wooConvNeXtV2Codesigning2023}), and Fig.\ref{f7i} (MobileNetV3\cite{howardSearching2019}), where the feature clusters form discrete and cohesive agglomerations.
Correlating these visual morphologies with our prior experimental data provides a compelling insight. Taking ResNet50 (linear distribution) and EfficientNetV2 (clustered distribution) as examples, the latter demonstrated superior recognition accuracy at low SNRs. Intuitively, a clustered distribution suggests that the model has learned more complex, non-linear relationships between features, resulting in more compact and independent feature clusters that are inherently more stable against noise. Conversely, a linear distribution may imply that the learned feature relationships are simpler and more susceptible to noise perturbations, which can disrupt cluster boundaries and thus degrade classification robustness.

In summary, the visual analysis of Fig.\ref{f7} leads to the following conclusions: 1) architectural integrity is crucial: AWSPNet's ability to identify targets and interference relies on its complete architecture; the front-end DTCWT or attention modules alone cannot effectively separate the signal features. 2) The backbone is foundational: the pre-trained backbone is the cornerstone of the model's performance, responsible for mapping the initially convoluted features into a highly separable space. 3) Feature morphology correlates with robustness: the distribution morphology of the final feature space exhibits a strong correlation with the model's noise robustness. Feature spaces characterized by clustered distributions (e.g., from EfficientNetV2) demonstrate superior robustness compared to those with linear distributions (e.g., from ResNet50), providing an important visual heuristic for understanding and selecting optimal network architectures.
\subsection{Target recognition and jamming suppression using AWSPNet}
\begin{figure*}[htbp]
	\centering
	\subfigure[]{
		\includegraphics[width=0.30\textwidth]{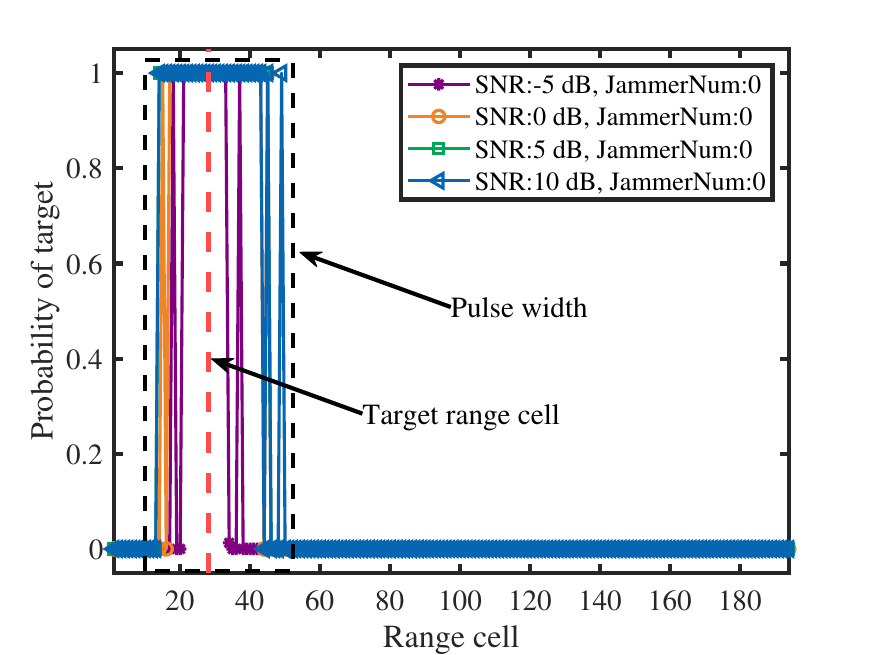}\label{f8a}
	}
	\,
	\subfigure[]{
		\includegraphics[width=0.30\textwidth]{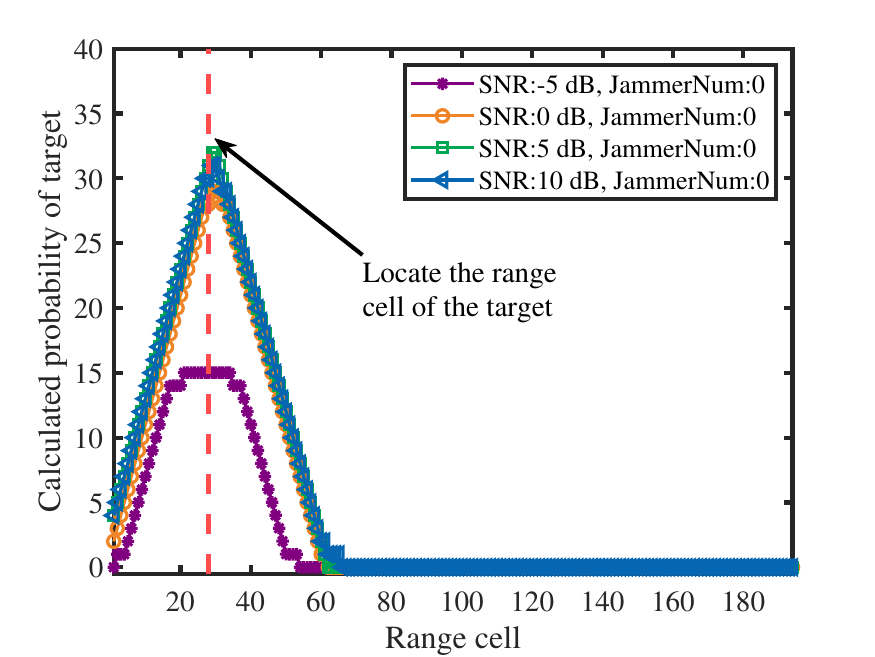}\label{f8b}
	}
	\,
	\subfigure[]{
		\includegraphics[width=0.30\textwidth]{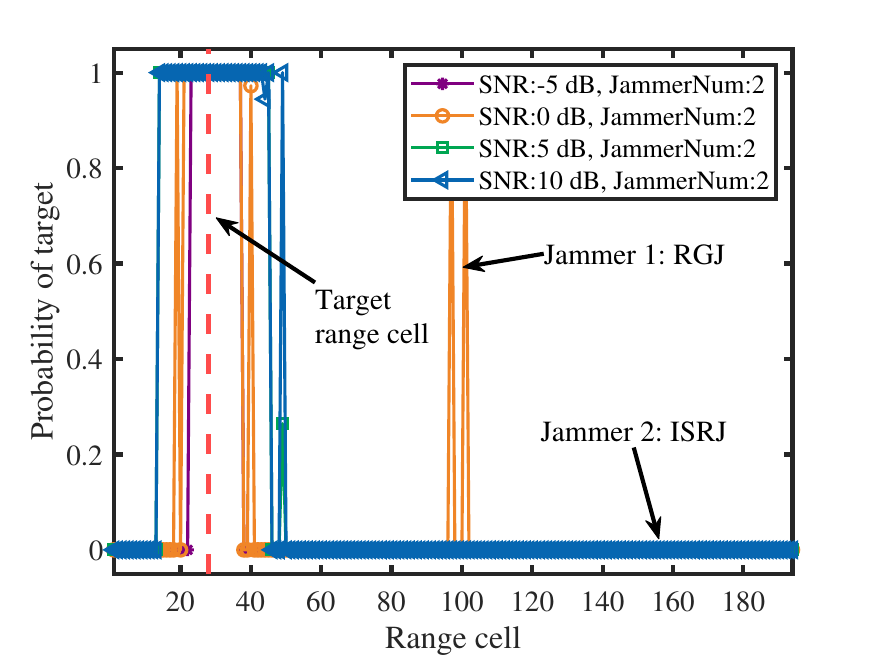}\label{f8c}
	}
	\,
	\subfigure[]{
		\includegraphics[width=0.30\textwidth]{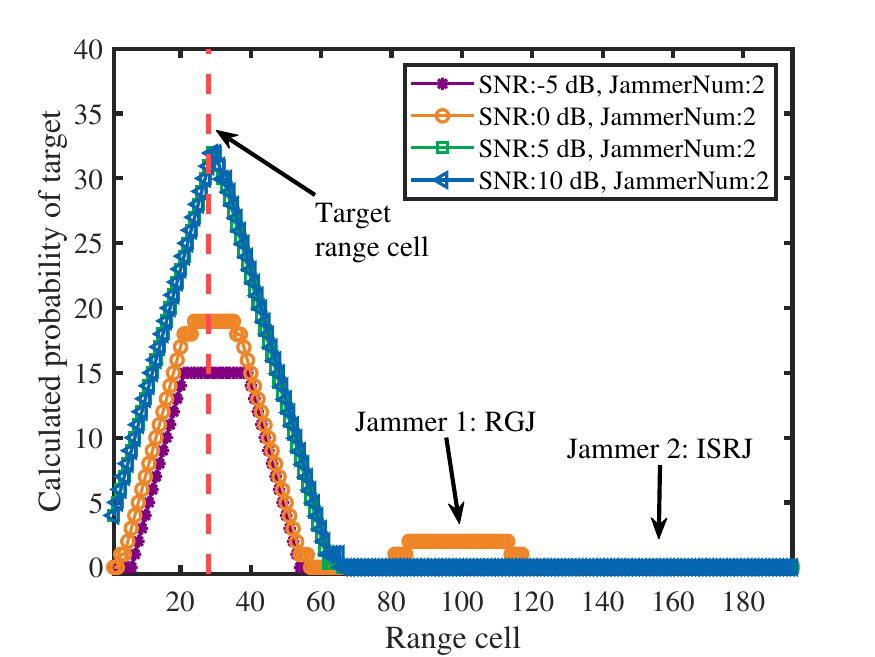}\label{f8d}
	}
	\,
	\subfigure[]{
		\includegraphics[width=0.30\textwidth]{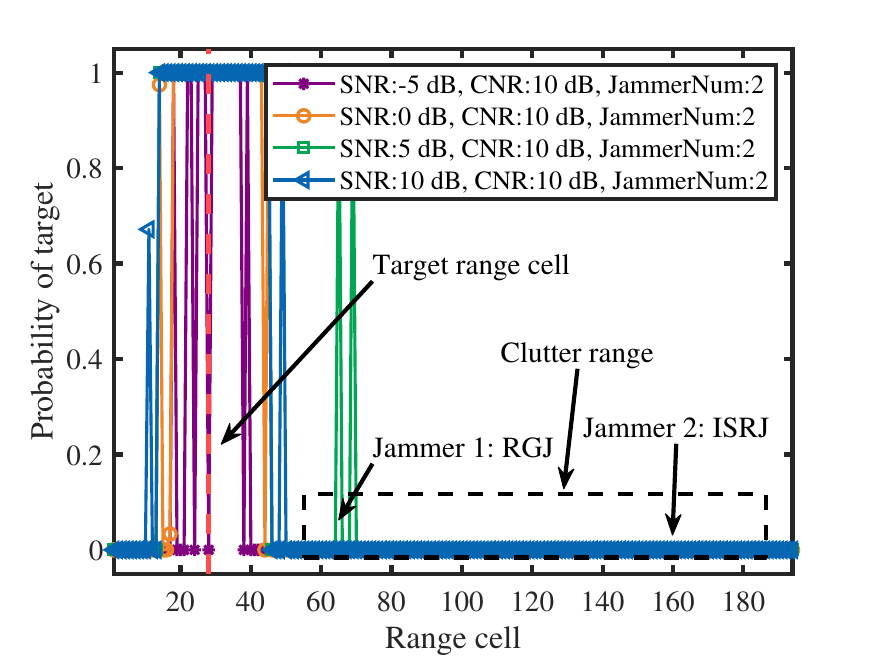}\label{f8e}
	}
	\,
	\subfigure[]{
		\includegraphics[width=0.30\textwidth]{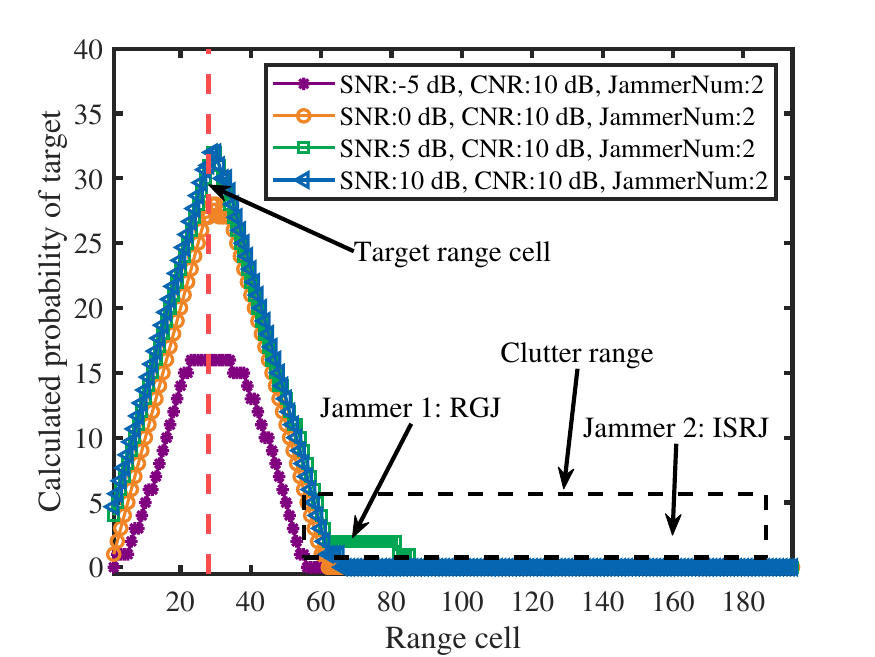}\label{f8f}
	}
	\caption{Target recognition and jamming suppression results of the AWSPNet (using EfficientnetV2 as pre-train net) model across different scenarios.
		(a) Direct recognition results for a single point target in an interference-free environment at SNRs of -5, 0, 5, and 10 dB.
		(b) Range detection results corresponding to a), obtained by convolving the output with a rectangular function equivalent in length to the pulse width. Peaks exceeding a set threshold indicate the target's range bin.
		(c) Recognition results for a single point target in the presence of two jammers at SNRs of -5, 0, 5, and 10 dB.
		(d) Corresponding range detection results for the two-interferer scenario in c).
		(e) Recognition results for a single point target with two interferers and a fixed CNR of 10 dB, evaluated across SNRs of -5, 0, 5, and 10 dB.
		(f) Corresponding range detection results for the high-CNR scenario in e). }
	\label{f8}
\end{figure*}
In this section, we investigate how the classification results from AWSPNet can be leveraged for jamming suppression. The core concept is to reframe the problem as a binary classification task: if we train the network to distinguish the target as one class and all other interferences as a second class, then the accurate identification of the target inherently achieves interference suppression. Following the procedure outlined in Algorithm \ref{a1}, we simulated the classification results for a single point target in the presence of various interferences and clutter, as shown in Fig.\ref{f8}.

First, all the results in Fig.\ref{f8} demonstrate that the proposed method can successfully locate the target. In scenarios with no interference or a limited number of interferers, as seen in Fig.\ref{f8a}, \ref{f8c}, and \ref{f8e}, the AWSPNet method effectively and continuously identifies the sequence of range bins occupied by the target. At higher SNRs, the recognition probability for these bins is 1.0, and their contiguous span corresponds to the length of the pulse width. As shown in Fig.\ref{f8b}, by subsequently convolving this probability output with a rectangular function equivalent to the pulse width (a process of probability accumulation), the central range bin of the target can be precisely located at the peak of the cumulative result.
Second, the method exhibits strong jamming suppression and resistance to false alarms. In Fig.\ref{f8c} and \ref{f8e}, the locations of interferers (such as RGJ and ISRJ) occasionally produce non-contiguous, isolated peaks with a probability of 1.0. These represent instances where the network misclassifies an interferer as the target in an individual range bin. However, due to their isolated and non-contiguous nature, these false alarm spikes are effectively smoothed and suppressed after the convolution with the rectangular function, as they fail to form a cumulative peak with sufficient energy. Consequently, the final detection results, shown in Fig.\ref{f8d} and \ref{f8f}, are free of false alarm targets caused by jammer, thereby achieving effective jamming suppression.
Third, the method demonstrates robustness at low SNRs. As shown by the purple curves (SNR = -5 dB) in Fig.\ref{f8a}, \ref{f8c}, and \ref{f8e}, the target recognition probability may not be a perfect, continuous block of high-probability bins due to the influence of noise, instead appearing intermittent. Despite this, the subsequent probability accumulation process effectively integrates these discontinuous recognition results into a distinct trapezoidal peak, as seen in Fig.\ref{f8b}, \ref{f8d}, and \ref{f8f}. The flat top of this peak clearly indicates the true range extent of the target, proving that the method can locate the target even when the initial recognition results are imperfect.
Finally, the method shows a generalization capability to jammer types not present in the training set, such as clutter. In the scenarios depicted in Fig.\ref{f8e} and \ref{f8f}, which include clutter (simulated with a clutter-to-noise ratio (CNR) equal to 10 dB), the model still accurately identifies and locates the target, even though clutter was not considered during network training. We attribute this to the significant differences in the time-frequency characteristics and probability distributions between the target and clutter. The model's powerful feature encoder is able to map the feature vectors of the target and clutter to points with sufficient separation in the high-dimensional space, enabling accurate target recognition even in the presence of this unseen jamming and showcasing generalization performance.

In summary, by combining the results of a binary target/non-target classification with a simple convolutional accumulation algorithm, the proposed AWSPNet framework not only identifies and locates targets but also effectively suppresses various known and unknown types of interference and clutter. It maintains robust performance at low SNRs, demonstrating its potential for target detection and jamming suppression in complex electromagnetic environments.

\section{Conclusion}
\label{s5}
This paper introduces the AWSPNet, a comprehensive deep learning framework designed to address the challenge of target recognition and jamming suppression for MIMO radar systems operating in complex electromagnetic environments. AWSPNet integrates the DTCWT for robust initial feature extraction, an attention mechanism for feature refinement, a pre-trained deep neural network for powerful representation learning, and a prototypical network for effective few-shot classification. Our objective was to create a model that not only achieves high recognition accuracy but also exhibits strong generalization, robustness to noise, and practical utility in a complete detection-to-suppression pipeline.
The experimental results presented validate the efficacy of the AWSPNet architecture. Ablation studies demonstrated that the integration of the DTCWT module, the attention mechanism, and transfer learning with pre-trained weights is fundamental to the model's high performance in low-SNR conditions (90.45\% accuracy at -6 SNR). Our analysis using t-SNE visualizations further illuminated the model's behavior, showing a clear correlation between the separability of feature clusters and classification accuracy across different noise levels, thus providing valuable insight into the network's internal mechanics. The visualizations confirmed that the complete, integrated architecture is necessary for effective feature separation.
Furthermore, we demonstrated the practical application of AWSPNet by embedding it within a time-domain sliding window algorithm. This approach enables the system to not only identify the presence and type of targets and interference but also to precisely locate them in time. By classifying segments of the signal as either target or non-target (interference/clutter), the framework effectively achieves jamming suppression. The system maintains robust performance even at low SNRs, highlighting its potential for deployment in real-world scenarios where distinguishing target echoes from powerful jamming is a significant challenge. 


%


\bibliographystyle{ieeetaes} 
\bibliography{Robust}

\begin{thebibliography}{10}
\providecommand{\url}[1]{#1}
\csname url@samestyle\endcsname
\renewcommand{\newblock}{\par}
\providecommand{\bibinfo}[2]{#2}
\providecommand{\BIBentrySTDinterwordspacing}{\spaceskip=0pt\relax}
\providecommand{\BIBentryALTinterwordstretchfactor}{4}
\providecommand{\BIBentryALTinterwordspacing}{\spaceskip=\fontdimen2\font plus
\BIBentryALTinterwordstretchfactor\fontdimen3\font minus
  \fontdimen4\font\relax}
\providecommand{\BIBforeignlanguage}[2]{{%
\expandafter\ifx\csname l@#1\endcsname\relax
\typeout{** WARNING: IEEEtran.bst: No hyphenation pattern has been}%
\typeout{** loaded for the language `#1'. Using the pattern for}%
\typeout{** the default language instead.}%
\else
\language=\csname l@#1\endcsname
\fi
#2}}
\providecommand{\BIBdecl}{\relax}
\BIBdecl

\bibitem{zhangJoint2025}
H.~Zhang, W.~Liu, Q.~Zhang, L.~Zhang, B.~Liu, and H.-X. Xu
\newblock  Joint power, bandwidth, and subchannel allocation in a uav-assisted
  dfrc network \newblock  \emph{IEEE Internet of Things Journal}, vol.~12,
  no.~9, pp. 11\,633--11\,651, May 2025.

\bibitem{zhangJoint2024}
H.~Zhang, W.~Liu, Q.~Zhang, and B.~Liu
\newblock  Joint customer assignment, power allocation, and subchannel
  allocation in a uav-based joint radar and communication network \newblock
  \emph{IEEE Internet of Things Journal}, vol.~11, no.~18, pp.
  29\,643--29\,660, Sep. 2024.

\bibitem{grecoCombined2005}
M.~Greco, F.~Gini, and A.~Farina
\newblock  Combined effect of phase and rgpo delay quantization on jamming
  signal spectrum \newblock  In \emph{IEEE International Radar Conference,
  2005.} Arlington, VA, USA: IEEE, Nov. 2005, pp. 37--42.

\bibitem{wangxuesongMathematic2007}
L.~J. WANG~XueSong
\newblock  Mathematic principles of interrupted-sampling repeater jamming
  (isrj) \newblock  \emph{Science in China Series F: Information Sciences},
  no.~1, pp. 113--123, 2007.

\bibitem{LiuImpact2019}
W.~Liu, J.~Meng, and L.~Zhou
\newblock  Impact analysis of drfm-based active jamming to radar detection
  efficiency \newblock  \emph{The Journal of Engineering}, vol. 2019, no.~20,
  pp. 6856--6858, Oct. 2019.

\bibitem{rongqingIntegrated2024}
W.~Rongqing, X.~I.~E. Jingyang, T.~Biao, X.~U. Shiyou, and C.~Zengping
\newblock  Integrated jamming perception and parameter estimation method for
  anti-interrupted sampling repeater jamming \newblock  \emph{Journal of
  Radars}, vol.~13, no.~6, pp. 1337--1354, Oct. 2024.

\bibitem{tianProduct2013}
X.~Tian, B.~Tang, and G.~{and Gui}
\newblock  Product spectrum matrix feature extraction and recognition of radar
  deception jamming \newblock  \emph{International Journal of Electronics},
  vol. 100, no.~12, pp. 1621--1629, Dec. 2013.

\bibitem{ruihuiIntelligent2024}
P.~Ruihui, W.~Xingrui, W.~Guohong, S.~Dianxing, Y.~Zhong, and L.~Hongwen
\newblock  Intelligent recognition and information extraction of radar complex
  jamming based on time-frequency features \newblock  \emph{Journal of Systems
  Engineering and Electronics}, vol.~35, no.~5, pp. 1148--1166, Aug. 2024.

\bibitem{zhouRecognition2023}
H.~Zhou, L.~Wang, and Z.~Guo
\newblock  Recognition of {{Radar Compound Jamming Based}} on {{Convolutional
  Neural Network}} \newblock  \emph{IEEE Transactions on Aerospace and
  Electronic Systems}, vol.~59, no.~6, pp. 7380--7394, Dec. 2023.

\bibitem{heDeep2016}
K.~He, X.~Zhang, S.~Ren, and J.~Sun
\newblock  Deep residual learning for image recognition \newblock  In
  \emph{2016 IEEE Conference on Computer Vision and Pattern Recognition
  (CVPR)}, Jun. 2016, pp. 770--778.

\bibitem{huangDensely2017}
G.~Huang, Z.~Liu, L.~Van Der~Maaten, and K.~Q. Weinberger
\newblock  Densely connected convolutional networks \newblock  In \emph{2017
  IEEE Conference on Computer Vision and Pattern Recognition (CVPR)}. Honolulu,
  HI: IEEE, Jul. 2017, pp. 2261--2269.

\bibitem{chenCompound2024}
H.~Chen \emph{et~al.}
\newblock  Compound {{Jamming Recognition Based}} on a {{Dual-Channel Neural
  Network}} and {{Feature Fusion}} \newblock  \emph{Remote Sensing}, vol.~16,
  no.~8, p. 1325, Jan. 2024.

\bibitem{zhouCompound2024}
H.~Zhou, L.~Wang, M.~Ma, and Z.~Guo
\newblock  Compound radar jamming recognition based on signal source separation
  \newblock  \emph{Signal Processing}, vol. 214, p. 109246, Jan. 2024.

\bibitem{lvMultilabel2024}
Q.~Lv, H.~Fan, J.~Liu, Y.~Zhao, M.~Xing, and Y.~Quan
\newblock  Multilabel {{Deep Learning-Based Lightweight Radar Compound Jamming
  Recognition Method}} \newblock  \emph{IEEE Transactions on Instrumentation
  and Measurement}, vol.~73, pp. 1--15, 2024.

\bibitem{quJRNet2020}
Q.~Qu, S.~Wei, S.~Liu, J.~Liang, and J.~Shi
\newblock  {{JRNet}}: Jamming recognition networks for radar compound
  suppression jamming signals \newblock  \emph{IEEE Transactions on Vehicular
  Technology}, vol.~69, no.~12, pp. 15\,035--15\,045, Dec. 2020.

\bibitem{liuPriorKnowledgeGuided2024}
Q.~Liu, X.~Zhang, and Y.~Liu
\newblock  A {{Prior-Knowledge-Guided Neural Network Based}} on {{Supervised
  Contrastive Learning}} for {{Radar HRRP Recognition}} \newblock  \emph{IEEE
  Transactions on Aerospace and Electronic Systems}, vol.~60, no.~3, pp.
  2854--2873, Jun. 2024.

\bibitem{wangApplication2024}
X.~Wang, S.~Chen, Y.~Zhu, S.~Zhang, X.~Li, and L.~Zhu
\newblock  Application of {{Wavelet Scattering Network}} and {{Ensemble
  Learning}} on {{Deception Jamming Recognition}} for {{Ultra-Wideband
  Detectors}} \newblock  \emph{IEEE Transactions on Microwave Theory and
  Techniques}, vol.~72, no.~4, pp. 2591--2601, Apr. 2024.

\bibitem{xiaoPSPNet2024}
S.~Xiao, S.~Zhang, M.~Jiang, and W.-Q. Wang
\newblock  Pspnet: Pretraining and self-supervised fine-tuning-based
  prototypical network for radar active deception jamming recognition with few
  shots \newblock  \emph{IEEE Geoscience and Remote Sensing Letters}, vol.~21,
  pp. 1--5, 2024.

\bibitem{shaDiffSwinT2023}
M.~Sha, D.~Wang, F.~Meng, W.~Wang, and Y.~Han
\newblock  Diff-{{SwinT}}: {{An Integrated Framework}} of {{Diffusion Model}}
  and {{Swin Transformer}} for {{Radar Jamming Recognition}} \newblock
  \emph{Future Internet}, vol.~15, no.~12, p. 374, Dec. 2023.

\bibitem{luoFewShot2023}
Z.~Luo, Y.~Cao, T.-S. Yeo, Y.~Wang, and F.~Wang
\newblock  Few-{{Shot Radar Jamming Recognition Network}} via {{Time-Frequency
  Self-Attention}} and {{Global Knowledge Distillation}} \newblock  \emph{IEEE
  Transactions on Geoscience and Remote Sensing}, vol.~61, no.~1, pp. 1--12,
  May 2023.

\bibitem{luoFewshot2024}
Z.~Luo, Y.~Cao, T.-S. Yeo, and F.~Wang
\newblock  Few-shot radar jamming recognition network via complete information
  mining \newblock  \emph{IEEE Transactions on Aerospace and Electronic
  Systems}, vol.~60, no.~3, pp. 3625--3638, Jun. 2024.

\bibitem{yangRadar2023}
Y.~Yang, Z.~Zhang, W.~Mao, Y.~Li, and C.~Lv
\newblock  Radar target recognition based on few-shot learning \newblock
  \emph{Multimedia Systems}, vol.~29, no.~5, pp. 2865--2875, Oct. 2023.

\bibitem{selesnickDualtree2005}
I.~Selesnick, R.~Baraniuk, and N.~Kingsbury
\newblock  The dual-tree complex wavelet transform \newblock  \emph{IEEE Signal
  Processing Magazine}, vol.~22, no.~6, pp. 123--151, Nov. 2005.

\bibitem{tanEfficientNetV22021}
M.~Tan and Q.~Le
\newblock  Efficientnetv2: Smaller models and faster training \newblock  In
  \emph{Proceedings of the 38th International Conference on Machine Learning}.
  Virtual Event: PMLR, Jul. 2021, pp. 10\,096--10\,106.

\bibitem{khoslaSupervised2021}
P.~Khosla \emph{et~al.}
\newblock  Supervised contrastive learning \newblock  In \emph{Proceedings of
  the 34th {{International Conference}} on {{Neural Information Processing
  Systems}}}, ser. {{NIPS}} '20. Red Hook, NY, USA: Curran Associates Inc.,
  Dec. 2020, pp. 18\,661--18\,673.

\bibitem{sparrowECM2006}
M.~J. Sparrow and J.~Cikalo
\newblock  Ecm techniques to counter pulse compression radar \newblock  US
  Patent US7\,081\,846B1, Jul., 2006.

\bibitem{wangInterference2023}
S.~Wang, J.~Du, W.~Fan, and F.~Zhou
\newblock  Interference suppression for synthetic aperture radar using
  dual-path residual network with attention mechanism \newblock  In
  \emph{IGARSS 2023 - 2023 IEEE International Geoscience and Remote Sensing
  Symposium}. Pasadena, CA, USA: IEEE, Jul. 2023, pp. 6787--6790.

\bibitem{cotterUses2020}
F.~Cotter
\newblock  Uses of complex wavelets in deep convolutional neural networks
  \newblock  Ph.D. dissertation, Department of Engineering, University of
  Cambridge, Cambridge, England, UK, Jun. 2020.

\bibitem{brunaInvariant2013}
J.~Bruna and S.~Mallat
\newblock  Invariant scattering convolution networks \newblock  \emph{IEEE
  Transactions on Pattern Analysis and Machine Intelligence}, vol.~35, no.~8,
  pp. 1872--1886, Aug. 2013.

\bibitem{zhangShuffleNet2017}
X.~Zhang, X.~Zhou, M.~Lin, and J.~Sun
\newblock  Shufflenet: An extremely efficient convolutional neural network for
  mobile devices \newblock  Dec. 2017.

\bibitem{howardSearching2019}
A.~Howard \emph{et~al.}
\newblock  Searching for mobilenetv3 \newblock  In \emph{2019 IEEE/CVF
  International Conference on Computer Vision (ICCV)}. Seoul, Korea (South):
  IEEE, Oct. 2019, pp. 1314--1324.

\bibitem{vandermaaten08a}
\BIBentryALTinterwordspacing
L.~{van der Maaten} and G.~Hinton
\newblock  Visualizing data using t-{{SNE}} \newblock  \emph{Journal of Machine
  Learning Research}, vol.~9, no.~86, pp. 2579--2605, Sep. 2008. [Online].
  Available: \url{http://jmlr.org/papers/v9/vandermaaten08a.html}
\BIBentrySTDinterwordspacing

\bibitem{wooConvNeXtV2Codesigning2023}
S.~Woo \emph{et~al.}
\newblock  {{ConvNeXt V2}}: {{Co-designing}} and scaling {{ConvNets}} with
  masked autoencoders \newblock  In \emph{2023 {{IEEE}}/{{CVF Conference}} on
  {{Computer Vision}} and {{Pattern Recognition}} ({{CVPR}})}. Vancouver, BC,
  Canada: IEEE, Jun. 2023, pp. 16\,133--16\,142.

\bibitem{xieAggregatedResidualTransformations2017}
S.~Xie, R.~Girshick, P.~Doll{\'a}r, Z.~Tu, and K.~He
\newblock  Aggregated residual transformations for deep neural networks
  \newblock  In \emph{2017 {{IEEE Conference}} on {{Computer Vision}} and
  {{Pattern Recognition}} ({{CVPR}})}. Honolulu, HI, USA: IEEE, Jul. 2017, pp.
  5987--5995.

\end{thebibliography}

\end{document}